%% file: main.tex
\RequirePackage{fix-cm}
\documentclass[twocolumn,epjc3]{svjour3}  
\RequirePackage[T1]{fontenc}
\usepackage[utf8]{inputenc}
\RequirePackage{graphicx}
\RequirePackage{mathptmx}      
\RequirePackage[numbers,sort&compress]{natbib}
\RequirePackage[colorlinks,citecolor=blue,urlcolor=blue,linkcolor=blue]{hyperref}
\input{basic/my_ds_macro.tex}
\usepackage{comment}
\usepackage{xcolor}
\usepackage{color}
\usepackage{adjustbox}
\usepackage{xurl}
\usepackage{tabularx}
\usepackage{float}
\usepackage{testhyphens}
\usepackage{hyphenat}
\usepackage{subfigure}
\usepackage{mathtools}
\input{basic/aria.def}
\usepackage{tabularx}
\usepackage{graphicx}
\input{basic/inst_dslm}
\usepackage{datetime2}

\usepackage[switch]{lineno}

\journalname{Eur. Phys. J. C}

\usepackage[normalem]{ulem}
\useunder{\uline}{\ul}{}
\makeatletter
\renewcommand{\thanksref}[1]{\nolinebreak\textsuperscript{\ref{#1}}\nolinebreak\checknextarg}
\newcommand{\checknextarg}{\@ifnextchar\bgroup{\nolinebreak\gobblenextarg}{}}
\newcommand{\gobblenextarg}[1]{ \textsuperscript{\nolinebreak\hspace{-4pt}\mbox{\nolinebreak$^,$\nolinebreak\ref{#1}\nolinebreak}\nolinebreak} \@ifnextchar\bgroup{\gobblenextarg}{}}
\begin{document}


\title{Measurement of isotopic  separation  of argon  with the prototype of the cryogenic distillation  plant  \Aria\ for dark matter searches}
\author{The DarkSide-20k Collaboration$^\text{\normalfont a,1}$}

\thankstext{e1}{e-mail: ds-ed@lngs.infn.it}

\institute{Please look at the end of the paper for the author list \label{addr1}}
%


\maketitle


\vskip 2cm 

\newcommand{\newtext}[1]{\textcolor{black}{#1}}

\renewcommand{\sout}[1]{\unskip}

\begin{abstract}
   \input{sections/abstract}
\end{abstract}

\input{sections/introduction}

\input{sections/simulation}

\input{sections/setup}

\input{sections/sampling}

\input{sections/hydraulic}
\input{sections/distillation}


\input{sections/multi}

\input{sections/comparison}

\input{sections/conclusions}
\input{sections/Acknoledgment}

\appendix
\input{sections/vola}


%



\bibliographystyle{basic/epj}


\input{basic/authors_ds20k-copy}

\end{document}

%% file: basic/my_ds_macro.tex
\usepackage[separate-uncertainty = true,multi-part-units=single]{siunitx}
\usepackage[version=4]{mhchem}

\usepackage{enumitem}
\setdescription{itemsep=0pt,parsep=0pt,labelsep=.2cm,leftmargin=2.2cm,labelwidth=2cm,listparindent=.7cm}
\setlist{nolistsep,leftmargin=1cm}
\newlist{enumcompactitem}{itemize}{3}
\setlist[enumcompactitem]{topsep=0pt,partopsep=0pt,itemsep=0pt,parsep=0pt}
\setlist[enumcompactitem,1]{label=\textbullet}
\setlist[enumcompactitem,2]{label=---}
\setlist[enumcompactitem,3]{label=*}
\newlist{enumcompactdesc}{description}{3}
\setlist[enumcompactdesc]{topsep=0pt,partopsep=0pt,itemsep=0pt,parsep=0pt}
\newlist{enumcompactenum}{enumerate}{3}
\setlist[enumcompactenum]{topsep=0pt,partopsep=0pt,itemsep=0pt,parsep=0pt}
\setlist[enumcompactenum,1]{label=\arabic*}
\setlist[enumcompactenum,2]{label=\alph*}
\setlist[enumcompactenum,3]{label=\roman*}
\DeclareSIUnit\c{\mbox{$c$}}
\DeclareSIUnit\magn{\mbox{$\times$}}
\DeclareSIUnit\min{min}
\DeclareSIUnit\week{week}
\DeclareSIUnit\month{mo}
\DeclareSIUnit\months{mos}
\DeclareSIUnit\year{yr}
\DeclareSIUnit\years{years}
\DeclareSIUnit\yr{yr}
\DeclareSIUnit\standard{std}
\DeclareSIUnit\str{sr}
\DeclareSIUnit\ppm{ppm}
\DeclareSIUnit\ppb{ppb}
\DeclareSIUnit\ppt{ppt}
\DeclareSIUnit\pe{PE}
\DeclareSIUnit\spe{SPE}
\DeclareSIUnit\pdm{PDM}
\DeclareSIUnit\ev{events}
\DeclareSIUnit\ct{counts}
\DeclareSIUnit\neutron{\mbox{$n$}}
\DeclareSIUnit\smp{samples}
\DeclareSIUnit\Sample{S}
\DeclareSIUnit\ch{ch}
\DeclareSIUnit\hit{hit}
\DeclareSIUnit\hits{hits}
\DeclareSIUnit\bin{(\mbox{5-PE}~bin)}
\DeclareSIUnit\sgm{\mbox{$\sigma$}}
\DeclareSIUnit\rms{RMS}
\DeclareSIUnit\keVee{\mbox{keV$_{e{\rm e}}$}}
\DeclareSIUnit\keVr{\mbox{keV$_{\rm nr}$}}
\DeclareSIUnit\eVee{\mbox{eV$_{\rm ee}$}}
\DeclareSIUnit\eVr{\mbox{eV$_{\rm nr}$}}
\DeclareSIUnit\ph{photon}
\DeclareSIUnit\el{\mbox{$e^-$}}
\DeclareSIUnit\pm{\mbox{PMT}}
\DeclareSIUnit\pixel{\mbox{pixel}}
\DeclareSIUnit\inch{''}
\DeclareSIUnit\foot{'}
\DeclareSIUnit\bit{bit}
\DeclareSIUnit\sample{samples}
\DeclareSIUnit\barn{barn}
\DeclareSIUnit\bara{bar}
\DeclareSIUnit\barg{barg}
\DeclareSIUnit\mlardepth{\mbox(meter~of~\LAr~depth)}
\DeclareSIUnit\Curie{Ci}
\DeclareSIUnit\psf{psf}
\DeclareSIUnit\pcf{pcf}
\DeclareSIUnit\parsec{pc}
\DeclareSIUnit\mwe{\mbox{m.w.e.}}
\DeclareSIUnit\liveday{\mbox{live-days}}
\DeclareSIUnit\days{\mbox{days}}
\DeclareSIUnit\miles{\mbox{miles}}
\DeclareSIUnit\lumens{\mbox{lm}}
\DeclareSIUnit\degreeC{\mbox{$^{\circ}$C}}
\DeclareSIUnit\degreeF{\mbox{$^{\circ}$F}}
\DeclareSIUnit\electron{\mbox{$e^-$}}
\DeclareSIUnit\Euro{\mbox{\euro}}
\DeclareSIUnit\cph{cph}
\DeclareSIUnit\neq{neq}
\DeclareSIUnit\normal{\mbox{N}}

\newcommand{\DSk}{\mbox{DarkSide-20k}}

\newcommand{\Aria}{\mbox{Aria}}

\newcommand{\LAr}{\ce{LAr}}

%% file: basic/inst_dslm.tex
\newcommand{\Alberta}{Department of Physics, University of Alberta, Edmonton, AB T6G 2R3, Canada}
\newcommand{\APC}{APC, Universit\'e de Paris, CNRS, Astroparticule et Cosmologie, Paris F-75013, France}
\newcommand{\AQLNGS}{INFN Laboratori Nazionali del Gran Sasso, Assergi (AQ) 67100, Italy}
\newcommand{\AQGSSI}{Gran Sasso Science Institute, L'Aquila 67100, Italy}

\newcommand{\AstroCeNT}{AstroCeNT, Nicolaus Copernicus Astronomical Center of the Polish Academy of Sciences, 00-614 Warsaw, Poland}
\newcommand{\Augustana}{Physics Department, Augustana University, Sioux Falls, SD 57197, USA}
\newcommand{\Belgorod}{Radiation Physics Laboratory, Belgorod National Research University, Belgorod 308007, Russia}

\newcommand{\BINP}{Budker Institute of Nuclear Physics, Novosibirsk 630090, Russia}
\newcommand{\Birmingham}{School of Physics and Astronomy, University of Birmingham, Edgbaston, B15 2TT, Birmingham, UK}
\newcommand{\BNLaddress}{Brookhaven National Laboratory, Upton, NY 11973, USA}
\newcommand{\BOINFN}{INFN Bologna, Bologna 40126, Italy}
\newcommand{\BOUniPHY}{Department of Physics and Astronomy, Universit\`a degli Studi di Bologna, Bologna 40126, Italy}
\newcommand{\CAUniCHE}{Department of Mechanical, Chemical, and Materials Engineering, Universit\`a degli Studi, Cagliari 09042, Italy}
\newcommand{\CAUniEEE}{Department of Electrical and Electronic Engineering, Universit\`a degli Studi di Cagliari, Cagliari 09123, Italy}
\newcommand{\CAUniPHY}{Physics Department, Universit\`a degli Studi di Cagliari, Cagliari 09042, Italy}
\newcommand{\CAINFN}{INFN Cagliari, Cagliari 09042, Italy}
\newcommand{\Carleton}{Department of Physics, Carleton University, Ottawa, ON K1S 5B6, Canada}

\newcommand{\CentroFermi}{Museo Storico della Fisica e Centro Studi e Ricerche Enrico Fermi, Roma 00184, Italy}
\newcommand{\CERNaddress}{CERN, European Organization for Nuclear Research 1211 Geneve 23, Switzerland, CERN}
\newcommand{\CIEMAT}{CIEMAT, Centro de Investigaciones Energ\'eticas, Medioambientales y Tecnol\'ogicas, Madrid 28040, Spain}

\newcommand{\CPPM}{Centre de Physique des Particules de Marseille, Aix Marseille Univ, CNRS/IN2P3, CPPM, Marseille, France}
\newcommand{\CTINFN}{INFN Catania, Catania 95121, Italy}
\newcommand{\CTUNI}{Universit\`a of Catania, Catania 95124, Italy}
\newcommand{\CTLNS}{INFN Laboratori Nazionali del Sud, Catania 95123, Italy}

\newcommand{\ENUniCEE}{Engineering and Architecture Faculty, Universit\`a di Enna Kore, Enna 94100, Italy}
\newcommand{\ETHZ}{Institute for Particle Physics, ETH Z\"urich, Z\"urich 8093, Switzerland}
\newcommand{\FNALaddress}{Fermi National Accelerator Laboratory, Batavia, IL 60510, USA}
\newcommand{\FortLewis}{Department of Physics and Engineering, Fort Lewis College, Durango, CO 81301, USA}
\newcommand{\GEUni}{Physics Department, Universit\`a degli Studi di Genova, Genova 16146, Italy}
\newcommand{\GEINFN}{INFN Genova, Genova 16146, Italy}
\newcommand{\Hawaii}{Department of Physics and Astronomy, University of Hawai'i, Honolulu, HI 96822, USA}
\newcommand{\Houston}{Department of Physics, University of Houston, Houston, TX 77204, USA}
\newcommand{\IHEPaddress}{Institute of High Energy Physics, Beijing 100049, China}

\newcommand{\JINR}{Joint Institute for Nuclear Research, Dubna 141980, Russia}
\newcommand{\Krakow}{M.~Smoluchowski Institute of Physics, Jagiellonian University, 30-348 Krakow, Poland}
\newcommand{\Kurchatov}{National Research Centre Kurchatov Institute, Moscow 123182, Russia}
\newcommand{\Laurentian}{Department of Physics and Astronomy, Laurentian University, Sudbury, ON P3E 2C6, Canada}
\newcommand{\Lancaster}{Physics Department, Lancaster University, Lancaster LA1 4YB, UK}
\newcommand{\Liverpool}{Department of Physics, University of Liverpool, The Oliver Lodge Laboratory, Liverpool L69 7ZE, UK}
\newcommand{\LNFINFN}{INFN Laboratori Nazionali di Frascati, Frascati 00044, Italy}
\newcommand{\LNLINFN}{INFN Laboratori Nazionali di Legnaro, Legnaro (Padova) 35020, Italy}
\newcommand{\Lodz}{Institute of Applied Radiation Chemistry, Lodz University of Technology, 93-590 Lodz, Poland}
\newcommand{\LPNHE}{LPNHE, CNRS/IN2P3, Sorbonne Universit\'e, Universit\'e Paris Diderot, Paris 75252, France}
\newcommand{\Mainz}{Institut f\"ur Kernphysik, Johannes Gutenberg-Universit\"at Mainz, D-55128 Mainz, Germany}
\newcommand{\Manchester}{Department of Physics and Astronomy, The University of Manchester, Manchester M13 9PL, UK}
\newcommand{\MEPhI}{National Research Nuclear University MEPhI, Moscow 115409, Russia}
\newcommand{\MendeleevUniverisity}{Mendeleev University of Chemical Technology, Moscow 125047, Russia}

\newcommand{\MIINFN}{INFN Milano, Milano 20133, Italy}
\newcommand{\MIPoliICA}{Civil and Environmental Engineering Department, Politecnico di Milano, Milano 20133, Italy}
\newcommand{\MIPoliCHE}{Chemistry, Materials and Chemical Engineering Department ``G.~Natta", Politecnico di Milano, Milano 20133, Italy}

\newcommand{\MIUni}{Physics Department, Universit\`a degli Studi di Milano, Milano 20133, Italy}
\newcommand{\MSU}{Skobeltsyn Institute of Nuclear Physics, Lomonosov Moscow State University, Moscow 119234, Russia}
\newcommand{\NAINFN}{INFN Napoli, Napoli 80126, Italy}
\newcommand{\NAUniPHY}{Physics Department, Universit\`a degli Studi ``Federico II'' di Napoli, Napoli 80126, Italy}
\newcommand{\NAUniCHE}{Chemical, Materials, and Industrial Production Engineering Department, Universit\`a degli Studi ``Federico II'' di Napoli, Napoli 80126, Italy}
\newcommand{\NAUniPHARM}{Pharmacy Department, Universit\`a degli Studi ``Federico II'' di Napoli, Napoli 80131, Italy}
\newcommand{\NAUniStruct}{Department of Strutture per l'Ingegneria e l'Architettura, Universit\`a degli Studi ``Federico II'' di Napoli, Napoli 80131, Italy}

\newcommand{\NSU}{Novosibirsk State University, Novosibirsk 630090, Russia}
\newcommand{\OACINAF}{INAF Osservatorio Astronomico di Capodimonte, 80131 Napoli, Italy}
\newcommand{\Petersburg}{Saint Petersburg Nuclear Physics Institute, Gatchina 188350, Russia}

\newcommand{\PIINFN}{INFN Pisa, Pisa 56127, Italy}
\newcommand{\PIUniPHY}{Physics Department, Universit\`a degli Studi di Pisa, Pisa 56127, Italy}
\newcommand{\PNNLaddress}{Pacific Northwest National Laboratory, Richland, WA 99352, USA}
\newcommand{\Princeton}{Physics Department, Princeton University, Princeton, NJ 08544, USA}
\newcommand{\Queens}{Department of Physics, Engineering Physics and Astronomy, Queen's University, Kingston, ON K7L 3N6, Canada}
\newcommand{\RHUL}{Department of Physics, Royal Holloway University of London, Egham TW20 0EX, UK}
\newcommand{\RMTreINFN}{INFN Roma Tre, Roma 00146, Italy}
\newcommand{\RMTreUni}{Mathematics and Physics Department, Universit\`a degli Studi Roma Tre, Roma 00146, Italy}
\newcommand{\RMUnoINFN}{INFN Sezione di Roma, Roma 00185, Italy}
\newcommand{\RMUnoUni}{Physics Department, Sapienza Universit\`a di Roma, Roma 00185, Italy}
\newcommand{\SAINFN}{INFN  Gruppo Collegato di Salerno, Sezione di Napoli, Salerno 84084, Italy}
\newcommand{\SAUni}{Physics Department, Universit\`a degli Studi di Salerno, Salerno 84084, Italy}
\newcommand{\SNOLABaddress}{SNOLAB, Lively, ON P3Y 1N2, Canada}

\newcommand{\STFCInterconnect}{Science \& Technology Facilities Council (STFC), Rutherford Appleton Laboratory, Technology, Harwell Oxford, Didcot OX11 0QX, UK}

\newcommand{\Temple}{Physics Department, Temple University, Philadelphia, PA 19122, USA}
\newcommand{\TNFBK}{Fondazione Bruno Kessler, Povo 38123, Italy}
\newcommand{\TNTIFPA}{Trento Institute for Fundamental Physics and Applications, Povo 38123, Italy}

\newcommand{\TOINFN}{INFN Torino, Torino 10125, Italy}
\newcommand{\TOPoli}{Department of Electronics and Communications, Politecnico di Torino, Torino 10129, Italy}

\newcommand{\TRIUMFaddress}{TRIUMF, 4004 Wesbrook Mall, Vancouver, BC V6T 2A3, Canada}

\newcommand{\UB}{Universiatat de Barcelona, Barcelona E-08028, Catalonia, Spain} 
\newcommand{\UCDavis}{Department of Physics, University of California, Davis, CA 95616, USA}
\newcommand{\UCRiverside}{Department of Physics and Astronomy, University of California, Riverside, CA 92507, USA}

\newcommand{\UCLA}{Physics and Astronomy Department, University of California, Los Angeles, CA 90095, USA}
\newcommand{\UCAS}{University of Chinese Academy of Sciences, Beijing 100049, China}
\newcommand{\UMass}{Amherst Center for Fundamental Interactions and Physics Department, University of Massachusetts, Amherst, MA 01003, USA}

\newcommand{\UnivAQ}{Universit\`a degli Studi dell’Aquila, L’Aquila 67100, Italy}
\newcommand{\UniversityofEdinburgh}{School of Physics and Astronomy, University of Edinburgh, Edinburgh EH9 3FD, UK}

\newcommand{\USP}{Instituto de F\'isica, Universidade de S\~ao Paulo, S\~ao Paulo 05508-090, Brazil}
\newcommand{\VTech}{Virginia Tech, Blacksburg, VA 24061, USA}
\newcommand{\Warwick}{University of Warwick, Department of Physics, Coventry CV47AL, UK}
\newcommand{\WilliamsCollege}{Williams College, Physics Department, Williamstown, MA 01267 USA}
\newcommand{\Zaragoza}{Centro de Astropart\'iculas y F\'isica de Altas Energ\'ias, Universidad de Zaragoza, Zaragoza 50009, Spain}
\newcommand{\ZaragozaARAID}{Fundaci\'on ARAID, Universidad de Zaragoza, Zaragoza 50009, Spain}
\newcommand{\CS}{CarboSulcis S.p.A. - Miniera Monte Sinni,  09010 Gonnesa, Italy}

%% file: sections/abstract.tex
The Aria cryogenic distillation plant, located in Sardinia, Italy, is a key component of the DarkSide-20k  experimental program  for WIMP dark matter searches at the INFN Laboratori Nazionali del Gran Sasso, Italy. Aria is designed to purify the argon, extracted from underground wells in Colorado, USA, and used as the DarkSide-20k target material,  to detector-grade quality. 

In this paper, we report the first measurement of argon isotopic separation by distillation with the 26~m tall Aria prototype. 
We discuss the  measurement of the operating parameters of the column 
and the observation of the simultaneous separation of the three stable argon isotopes: \ce{^36Ar}, \ce{^38Ar}, and \ce{^40Ar}. 
We also provide a detailed comparison of the experimental results with commercial process simulation software.

This measurement of isotopic separation of argon is a significant achievement for the project, building on the success of the initial demonstration of isotopic separation of nitrogen using the same equipment in 2019.

%% file: sections/introduction.tex
\section{Introduction}
\label{sec:intro}
Aria  is an industrial-scale plant comprising a $350$~m cryogenic  distillation column for isotopic separation~\cite{Darkside:2021}, the tal\-lest ever built, and  is currently being installed in a mine shaft at Carbosulcis S.p.A., Nuraxi-Figus (SU), Italy. 
Aria is designed to reduce the isotopic abundance of \ce{^39Ar}  in the low-radioacti\-vity argon, or undergorund argon (UAr), extracted from underground sources and used for  dark matter searches. 

A full description of the Aria plant and of the column structure and performance can be found in Ref.~\cite{Darkside:2021}.

In recent years, interest in UAr has grown to include its use in searches for neutrino-less double-beta decay with the LEGEND experiment~\cite{Legend}; in the measurement of coherent elastic neutrino-nucleus scattering with the COHERENT experiment~\cite{Coherent}); and in measuring low-energy processes with one of the planned   modules \ of the DU\-NE ex\-pe\-riment  (\href{https://doi.org/10.48550/arXiv.2203.08821}{PN\-NL-\-SA-171088}). The projected demand for UAr varies great\-ly, from one tonne for the COHERENT experiment to several thousand tonnes for a detector of similar size to DUNE. The required \ce{^39Ar} depletion factor with respect to atmospheric argon levels spans from a factor $1000$ for COHERENT and DarkSide-20k to more than  $10000$ for future  dark matter searches with, {\it e.g.}, a tonne-scale DarkSide-LowMass experiment (\href{https://doi.org/10.48550/arXiv.2209.01177}{arXiv:2209.01177}). 

The first application  of the Aria plant will be to purify 120~t of UAr through cryogenic distillation, which will be used in the \DSk\ experiment~\cite{Aalseth:2018gq}. However, the high depletion factors required for future projects also demonstrate the need for isotopic separation using Aria.

The initial phase of the Aria program includes assessing the efficiency of the distillation process and determining the thermodynamic  parameters of the prototype distillation plant.
A successful nitrogen isotope separation campaign  was conducted with the same plant in 2019 and described in Ref.~\cite{Darkside:2021}. The prototype plant is a shortened version of the Aria column, consisting of a top condenser, a bottom reboiler, and one central module (out of $28$). It is equipped with all the auxiliary instrumentation of the full column and located in a surface building at Carbosulcis S.p.A.
The results obtained using nitrogen were very promising. However, argon has distinct thermodynamic properties, such as its boiling point in relation to pressure and its latent heat at the boiling point, compared to nitrogen. Therefore, a separate characterization of the plant's performance is necessary when using argon.

This article presents the results of a distillation run for the  separation of argon isotopes conducted  in 2021. Whereas \ce{^39Ar} is only present in traces,
the two stable isotopes other than \ce{^40Ar} in atmospheric argon, \ce{^36Ar} and \ce{^38Ar}, have a non-negligible isotopic abundance of 0.334\% and 0.063\%, respectively \cite{Lee,Bohlke}, and can be used to characterize the distillation performance of the plant. 

With respect to Ref.~\cite{Darkside:2021}, this paper presents a more  in-depth characterization of the plant performance, including hydraulic parameters, details of the distillation process, and a thorough comparison between the actual performance and that obtained using a commercial process simulation software. 
We have also demonstrated through experimentation that multi-component distillation modeling is not required for argon distillation, despite the presence of more than two isotopes in the input feed. This  was already  hypothesized  in Ref.~\cite{Darkside:2021} without experimental proof.
Here we briefly reproduce a few basic formulae of distillation columns.
For a two-component distillation at total reflux,  i.e. without a significant amount of argon entering or exiting the column during distillation, the separation between isotopes $i$ and $j$ between the top (T) and the bottom (B) of the column is given by:

\begin{equation} 
S^{TB}_{i-j}=(\alpha_{i-j})^N,\label{eq1}
\end{equation}

where $\alpha_{i-j}$ is the relative volatility between the lighter, more volatile isotope~$i$ and the heavier, less volatile isotope~$j$,  and 
$N$ is the  number of theoretical stages. The  separation can be measured from:

\begin{equation}
\label{frazione}
S^{TB}_{i-j}=\frac{(R_{i-j})_T}{(R_{i-j})_B},
\end{equation}
where $(R_{i-j})_{T(B)} $  is the isotopic ratio, i.e. the relative abundance between the lighter and the heavier isotope, measured at the top (bottom) of  the column. 

The logarithm of the relative volatility, also known as the Vapor Pressure Isotope Effect (VPIE), is given by:

\begin{equation} \label{nuova}
\mathrm{VPIE}=\ln{\alpha_{i-j}}\simeq 
\ln\frac{{P_i}}{{P_j}} \simeq \ln\frac{{f_c}}{{f_g}}, \end{equation}

with $P_i$ and $P_j$  the  vapor pressures of the two isotopes $i$ and $j$ and $f_c$ and $f_g$ the reduced partition functions in the condensed phase and in the ideal gas.

In a column with structured packing like Aria,  the distillation performance is related to the height equivalent to a theoretical plate or stage, HETP, and is given by:


\begin{equation} \mathrm{HETP}=\frac{h}{N} \label{eq2}
\end{equation}

where $h$ is the cumulative height of the packing material of the column.  Combining  Eq.~\ref{eq1} and Eq.~\ref{eq2}, it follows that:

\begin{equation} \mathrm{HETP}={\frac{h \times \ln(\alpha_{i-j})}{\ln(S^{TB}_{i-j})}}
\label{eq3}
\end{equation}

The rest of the paper is organized as follows.  In Sect.~\ref{sec:status} we discuss the simulation programs  used in this work.  In Sect.~\ref{exp} we present the experimental setup. In Sect.~\ref{hydraulic} we discuss the pressure drop along the column and the liquid holdup. In Sect.~\ref{Results} and~\ref{3otto} we discuss the distillation measurements and in Sect.~\ref{sec:comp} we compare them with the simulations. In Sect.~\ref{sec:conclusions} we conclude and give some perspectives. In~\ref{vola}
we derive an estimate of the relative volatilities  of \ce{^40Ar} with respect to \ce{^36Ar} and \ce{^38Ar} from existing measurements.

%% file: sections/simulation.tex
\section{The simulation programs}
\label{sec:status}
We use the Aspen HYSYS (\textcopyright 2022 Aspen Technology Inc) software, a powerful tool for industrial process simulations, to model our system. 
HYSYS performs rigorous distillation calculations, i.e.  solve the equations for mass, component, energy, and equilibrium balance for each stage of the  column. 
Rigorous calculations also allow the simulation of multi-component distillation processes, with three or more components modeled simultaneously. 
The standard HYSYS library of chemicals does not include isotopes. For this simulation, we added the argon isotopes to the HYSYS library, together with their molecular weight and Antoine equation parameters.

The results for binary distillation obtained with HYSYS were compared with those from a calculation with the graphical McCabe-Thiele (MCT) method.
%
%



We also used the Sulcol\ce{^T^M} $3.5$ (Sulzer) software package for structured and random packing design to model the hydraulic parameters of the distillation column.
Taking input properties, such as the density and viscosity of the gas and liquid phases, the surface tension, and the mass flow rates of the gas and the liquid inside the column, Sulcol calculates column parameters including the pressure drop per unit length and  the liquid hold-up of the column.

%% file: sections/setup.tex
\section{The experimental setup and measurements}
\label{exp}
Before starting operations, 
the  column and all the process lines were evacuated through a scroll pump. 
The column was then filled with argon gas. 
The total amount of argon filling the column 
was about $250$ kg.

Plant operation  followed a  procedure similar to our previous run with nitrogen~\cite{Darkside:2021}. 
Automatic process control  was implemented for most 
of the system.
For this run, it took $12$~days of operation to stabilize the plant and start the distillation measurements reported below, which lasted about 4 days.
As for the past run, nitrogen was used as the refrigerant fluid.

The distillation was performed in total reflux mode.  Abo\-ut 1~L~h$^{-1}$ of gas was extracted for the sampling, corresponding to an overall loss during the run of $<$0.1\% of the loaded argon mass. Therefore,  we expect Eq.~\ref{eq1} to hold.

\begin{figure}
\centering
\subfigure[]{
\includegraphics[width=\columnwidth]{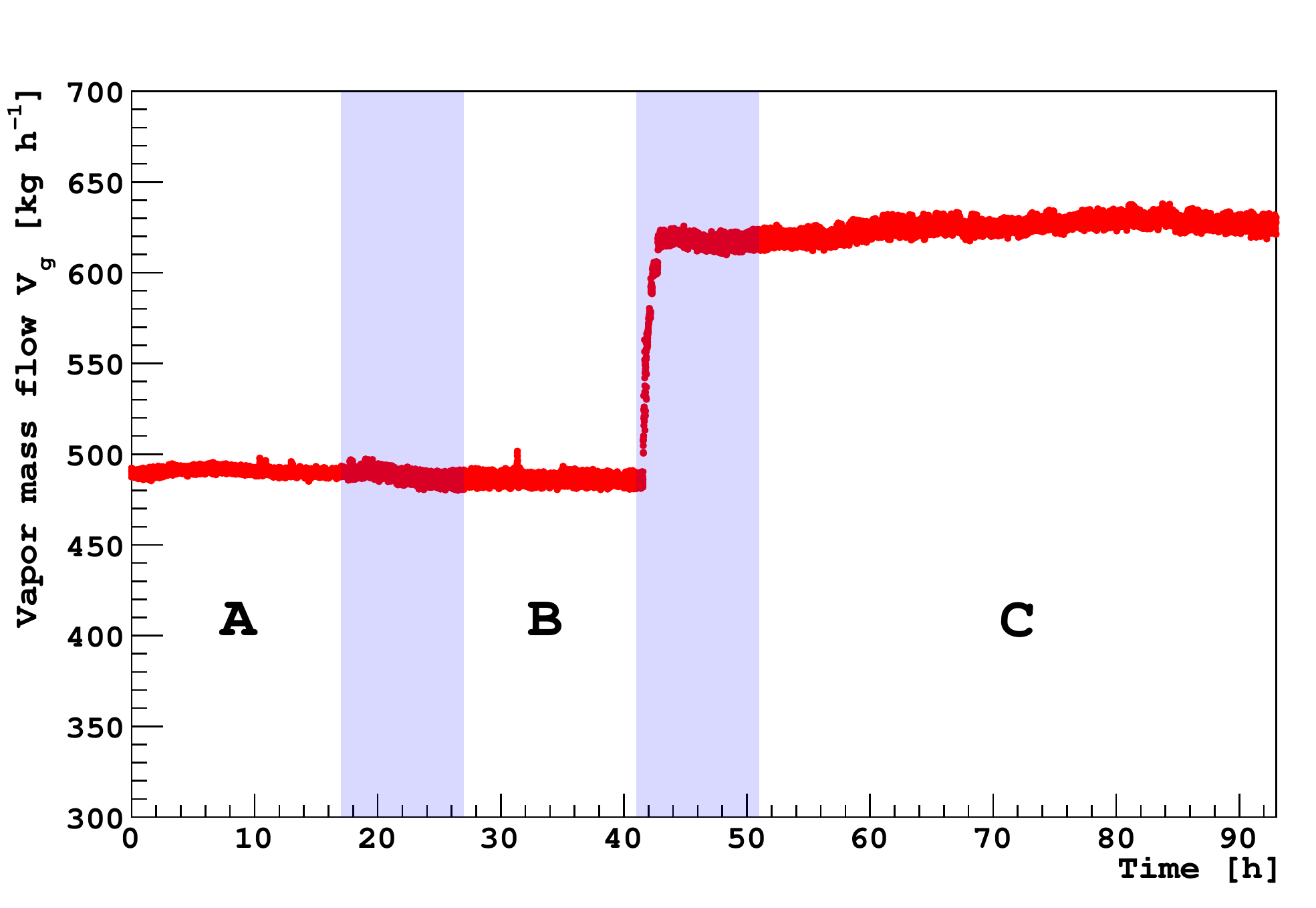}
\label{totalissimoa}}
\subfigure[]{
    \includegraphics[width=\columnwidth]{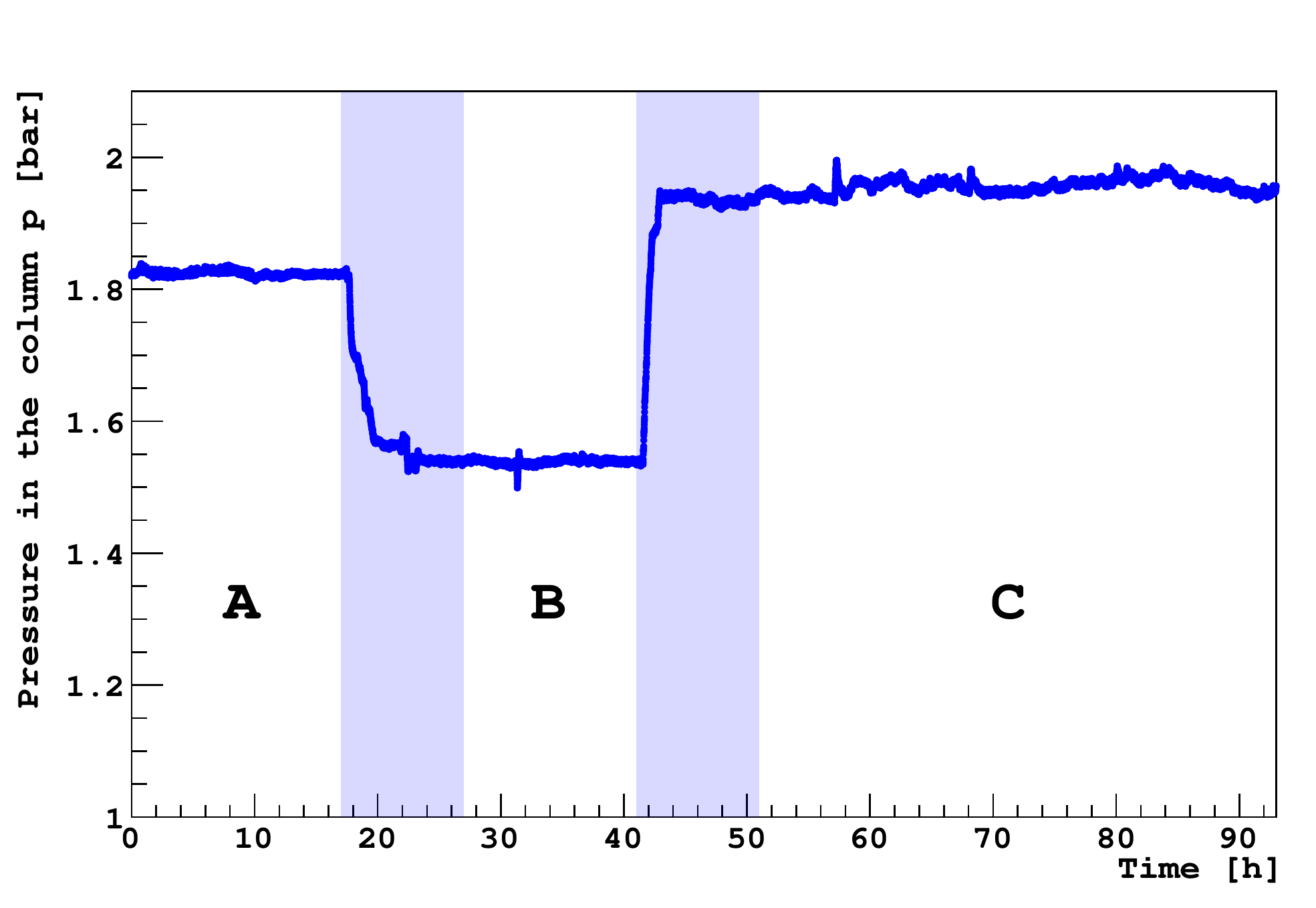}
    \label{totalissimob}
}
\caption{Vapor mass flow rate of  nitrogen in the auxiliary system (red, \ref{totalissimoa}) and average pressure inside the column $p$ (blue, \ref{totalissimob}) as a function of time. The shaded regions  correspond to periods with changing conditions in the column and are not included in  the data analysis.}
\label{totalissimo}
\end{figure}

The operating conditions of the argon distillation runs are presented in Fig.~\ref{totalissimo}.
The top panel (Fig.~\ref{totalissimoa}) displays 
the vapor mass flow rate of nitrogen in the auxiliary system (red); the bottom panel (Fig.~\ref{totalissimob}) shows the average pressure $p$ inside the column (blue) as a function of time. The time averages of the same quantities are summarized in  Table~\ref{tab_dati}, together with the maximum pressure variation during the run, $\delta_{p}$. 
The collected data is divided in three runs (A, B, and C) corresponding to changing operating conditions. 
At the beginning of run~A, the vapor mass flow rate of nitrogen was about 10\% lower than the plant design value of 550~kg~h$^{-1}$.
In run~B, we kept the vapor mass flow rate unchanged and lowered the pressure inside the column by increasing the amount of nitrogen introduced into the auxiliary system.
In run~C we increased the vapor mass flow rate to a value about~10\% higher than the design value.
The lowest average pressure achieved inside the column was about~18\% larger than the design value described in Ref.~\cite{Darkside:2021}.
With  argon distillation, we have achieved a ten-fold lower pressure variation during the run, than during the  nitrogen distillation run discussed in Ref.~\cite{Darkside:2021}.

\begin{table}
\setlength{\tabcolsep}{12pt}
\centering
\caption{Time averaged operational  parameters during runs A, B and C: 
 pressure inside the column, maximum pressure variation $\delta_{p}$ during the run, and vapor mass flow rate of nitrogen in the auxiliary system.}
\begin{tabular}{@{} l c c c @{}}
\hline\noalign{\smallskip}
&$p$ (bar) &$\delta_{p}$ (bar)&  vapor flow rate (kg h$^{-1}$) \\
\noalign{\smallskip}\hline\noalign{\smallskip}
A&$1.824\pm 0.001$&$\pm 0.02$&$491\pm 1$ \\ 
B&$1.544\pm 0.001$&$\pm 0.02$&$487\pm 1$\\ 
C&$1.953\pm 0.001$&$\pm 0.04$&$624\pm 1$ \\ 
\hline
\end{tabular}
\label{tab_dati}
\end{table}
Table~\ref{tab_altridati} shows three derived operational parameters: the time-averaged  saturation temperature, $T$, obtained from the Antoine equation using the measured pressures from Table~\ref{tab_dati}; 
the argon vapor mass flow rate, V, derived from the energy balance in the reboiler; and the thermal power of the process, given by the product of nitrogen  latent heat and  nitrogen mass flow rate.

\begin{table}
\setlength{\tabcolsep}{20pt}
\centering
\caption{Derived operational parameters: time-averaged  saturation temperature, $T$,  argon vapor mass flow rate, $V$, and the thermal power of the process, $Q$.}
\begin{tabular}{@{} l c c c @{}}
\hline\noalign{\smallskip}
&$T$ (K)  & V (kg h$^{-1}$) & $Q$ (kW)  \\
\noalign{\smallskip}\hline\noalign{\smallskip}
A&$93.3\pm 0.1$&$543\pm 1$&$23.0\pm 0.1$ \\ 
B&$91.5\pm 0.1$&$544\pm 1$&$23.8\pm 0.1$\\
C&$94.0\pm 0.1$&$689\pm 1$&$29.7\pm 0.1$\\
\hline
\end{tabular}
\label{tab_altridati}
\end{table}

%% file: sections/sampling.tex
\subsection{The gas sampling system}

\begin{figure*}
\centering
\includegraphics[width=\textwidth]{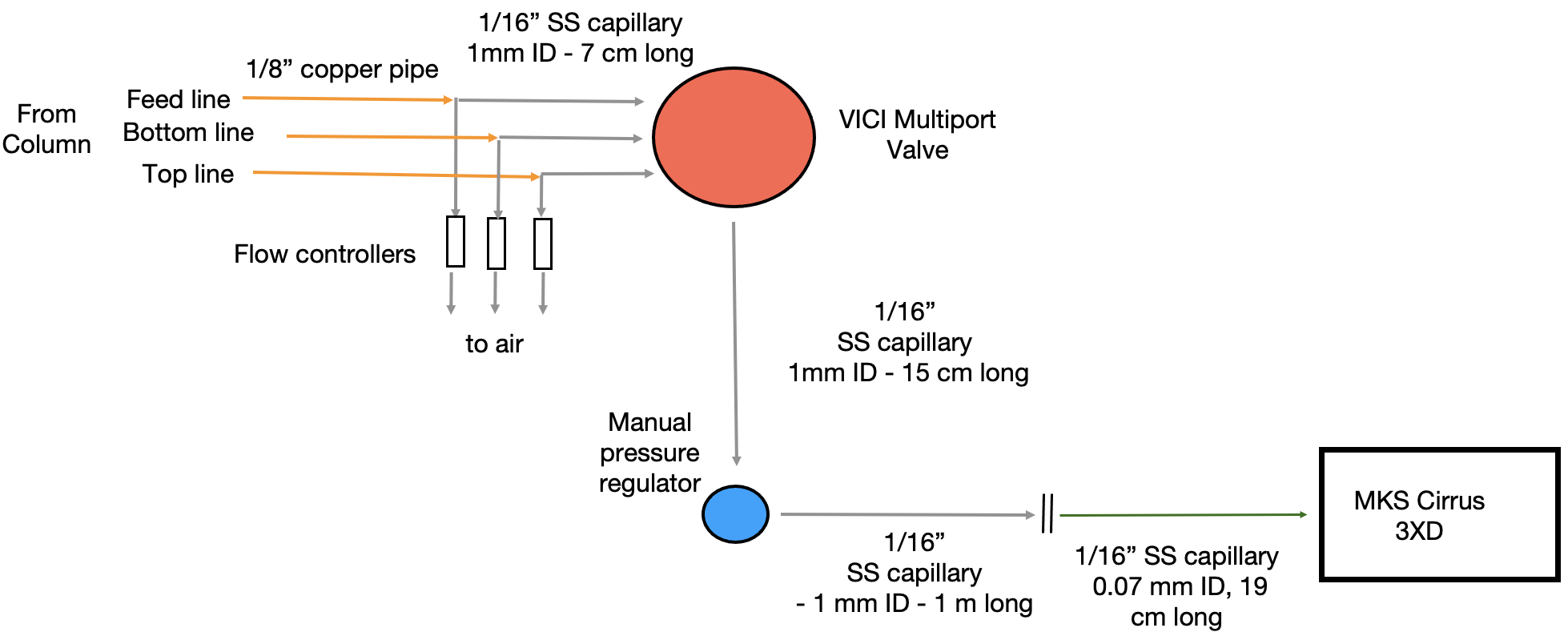}
\caption{Schematic of the gas sampling system. }
\label{sampling} 
\end{figure*}

\begin{figure}
\centering
\includegraphics[width=\columnwidth]{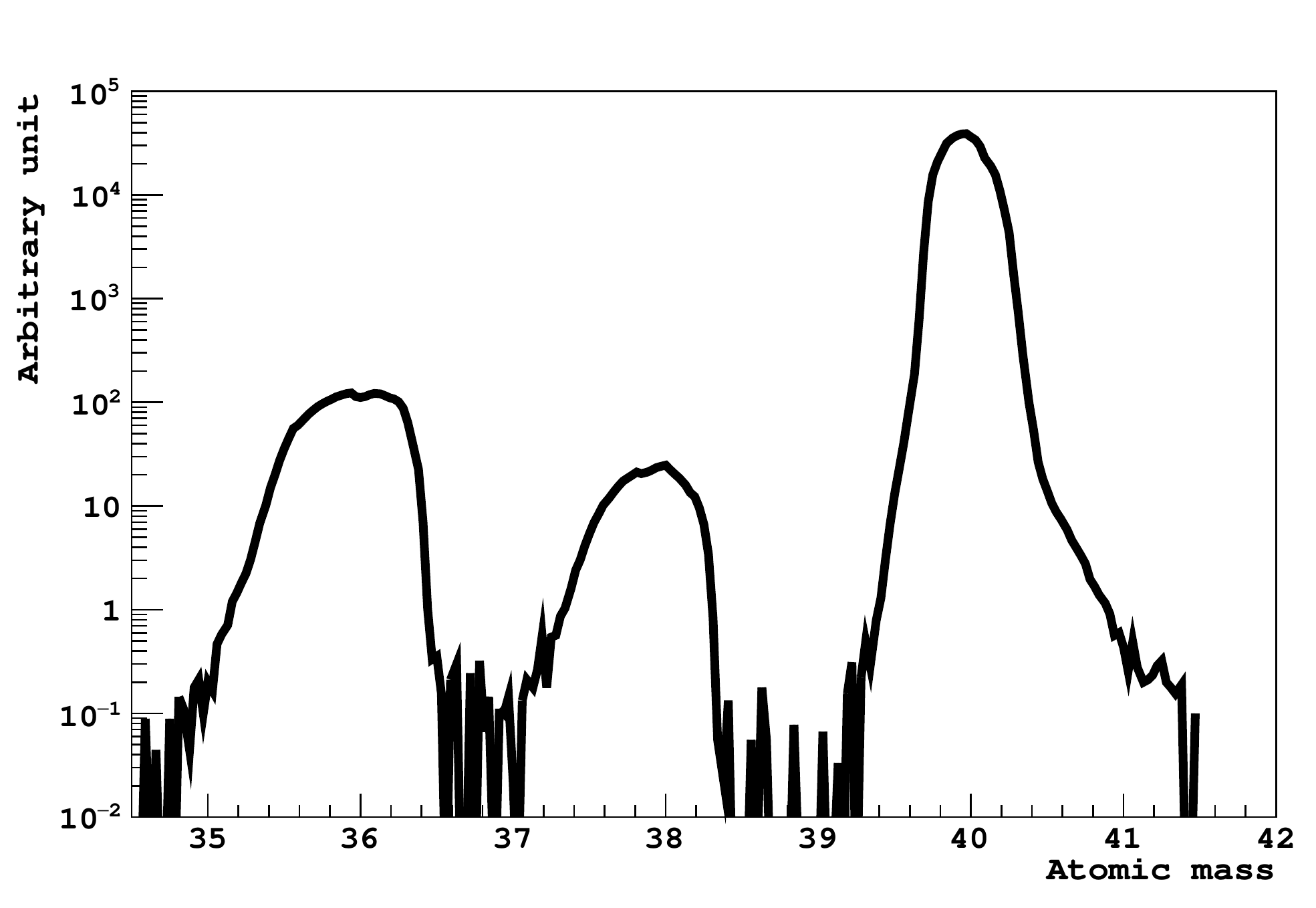}
\caption{Example of a background subtracted mass spectrum measured during the run, zoomed in the region of interest for argon isotopes. }
\label{spettro} 
\end{figure}

%
The instrument used to measure the isotopic distillation performance is an MKS Instruments, Cirrus\texttrademark\  3-XD quadrupole mass spectrometer. 
During the run, the mass spectrometer was continuously in operation to sample the gas coming from the top (Top) and bottom (Bottom) of the column, and from the gas bottles (Feed), as shown in Fig.~\ref{sampling}. 
The gas from the column flows through~1/8"   copper lines between 4~m (Bottom) and 30~m (Top) in length. These lines are  split before the mass spectrometer. 
Most of the gas flow, controlled via mass flow controllers set around $50$~mL min$^{-1}$, purge the lines and is vented to air. The gas along the other branch flows through, respectively, a ~1/16"~($1$ mm ID) stainless steel (SS), $7$~cm long, capillary line; a multi-port valve (VALCO Instruments); another 1/16"~($1$~mm ID), $15$~cm long SS capillary line; and a 
  manual pressure regulator, to limit the gas pressure to a maximum of $1$ bar, thus protecting the input of the mass spectrometer.
The gas then passes through a 1/16"~($1$~mm ID), $1$~m long SS line,   and a $19$~cm ($0.07$~mm ID) long SS capillary, to the mass spectrometer.
The inlet flow  of the mass spectrometer is  $20$~mL min$^{-1}$. 
Before the run, the response time of the sampling system was determined by connecting a bottle of argon and one of nitrogen to the multi-inlet valve. It took about 
 $10$~min to see a change in the measured  gas composition and 40-45~min to see it stabilize. This lag time is caused by the combination of  the relatively large buffer volume of the  manual pressure regulator and the low gas  inlet flow.
During gas sampling of argon from the column, the inlet valve was programmed to switch among the three input capillaries every hour. This is a compromise between the response time of the sampling system and the need for quickly detecting changes in isotopic mass fraction in the column when the operating parameters are changed.  
In Fig.~\ref{spettro} we show an example of a background subtracted mass spectrum from the spectrometer during the run, zoomed in the region of interest for argon isotopes 
and featuring well-separated peaks. 

\begin{figure}
\centering
\includegraphics[width=\columnwidth]{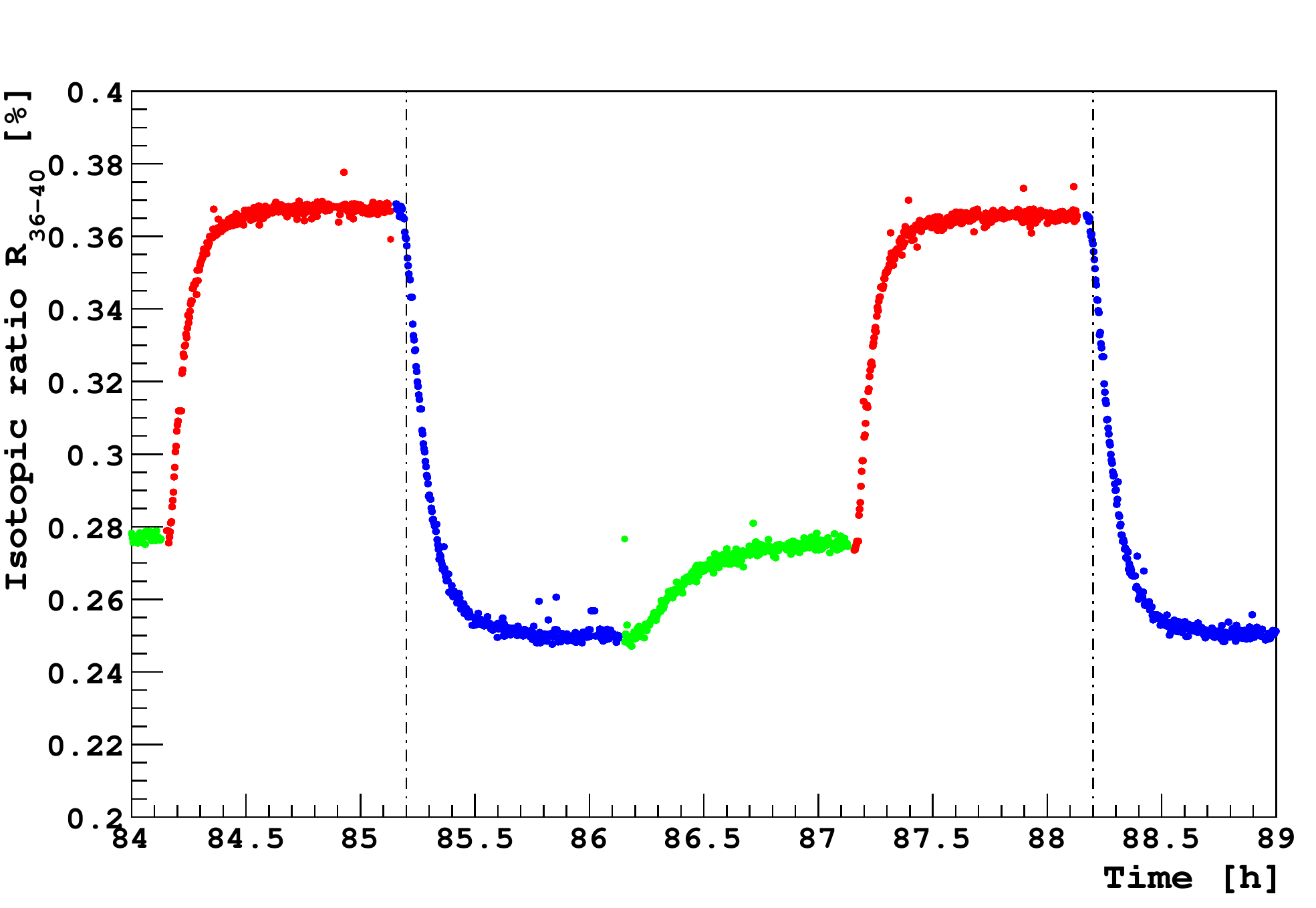}
\caption{Isotopic ratio, $R_{36-40}$,  as a function of time, for a time slice of run C. Every hour, the inlet valve is switched from Top (red) to Bottom (blue) and Feed (green).
 The isotopic ratios were calculated at the time when the inlet valve is switched to the Bottom sampling (e.g. at times marked by dashed vertical lines), and 
 only considering  the last 15 min before each switching of the inlet valve. 
}
\label{Setup23}
\end{figure}

We obtain the isotopic ratios $R_{36-40}$ and $R_{38-40}$ as ratios of the \ce{^36Ar} (\ce{^38Ar}) and \ce{^40Ar} peak heights. 
$R_{36-40}$ is displayed vs time in Fig.~\ref{Setup23},
where each point corresponds to a measurement derived from a spectrum like the one in Fig.~\ref{spettro}. 
Fig.~\ref{Setup23} clearly shows that the measurement of the isotopic ratio settles to a plateau in approximately 45 min.  
Because of this delayed response, we only considered the last 15~min of each 1-hour sampling period during the measurement of the distillation performance,.
The isotopic ratios were calculated with the inlet valve switched to the Bottom sampling (e.g. at times marked by dashed vertical lines). We estimated a systematic uncertainty of the order of 2\% on the measured isotopic ratio due to this choice, by comparing it with other possible choices of times for calculating the isotopic ratios.

%% file: sections/hydraulic.tex



\section{Pressure drop along the column and liquid holdup}
\label{hydraulic}
%



The Aria column is filled with structured stainless
steel packing (Sulzer CY gauze) sections  \AriaColumnModuleSectionHeight\ long, four per module, interleaved with a liquid
distributor to optimize flow uniformity across the column cross-section, whose structure is discussed below.
The purpose of the packing is to  support the liquid phase, for optimal thermodynamic contact between the rising vapor and the descending liquid. During operations, the liquid is dispersed as a film coating the surfaces of the packing and immersed in the vapor phase. This configuration is referred to as irrigated bed. 

Two very relevant operational parameters of the packed distillation column are the pressure drop per unit length and the liquid holdup~\cite{Kister:1992}, described hereafter. 

\subsection{Pressure drop}
\label{sec:Pressure drop}
The pressure drop is due to friction and directional changes in the gas flow inside the column caused by the packing and the distributors.
Therefore, the total pressure drop in the column, $\Delta p_{tot}$, is given by the sum of the pressure drop in the distributors, $\Delta p_{\mathrm{distributors}}$, and in the wetted packing elements, $\Delta p_{\mathrm{packing,irr}}$ as:

\begin{equation}
\Delta p_{tot}=\Delta p_{\mathrm{distributors}}+\Delta p_{\mathrm{packing,irr}} 
\end{equation}

\subsubsection{Calculation of the pressure drop due to the distributors}

The distributors, whose picture can be found in Ref.~\cite{Darkside:2021}, are made by horizontal plates, intercepting the  downward liquid argon flow along the column. The plates are perforated, with 10~cm hollow vertical pipes or chimneys with a top cap at the hole locations on the upper side, which are uniformly distributed on the plate surface. The vapor flows up through these chimneys.
The liquid formed on the distributor plate is streamed, through 0.3~cm holes located at 3~cm, 4~cm, and 5~cm height in  the chimneys, to the packing section below.  

In our calculations, the distributed pressure drop due to the short pipes can be neglected and the distributor is, therefore, approximated 
as a perforated sheet with holes. 
The pressure drop from the four distributors, $\Delta p_{\mathrm{distributors}}$, is determined following Ref.~\cite{caleffi28}:

\begin{equation}
\Delta p_{\mathrm{distributors}}= 4\times \zeta \; \rho{_\mathrm{G}}\; \frac{v_\mathrm{G}^2}{2}, 
\label{deltap}
\end{equation}

where $\zeta$ is the concentrated loss coefficient, and $\rho{_\mathrm{G}}$ and $v_\mathrm{G}$ are the density and the velocity of the argon gas, respectively. 
The value of $\zeta$ is evaluated by interpolating the data reported on page~18 of Ref.~\cite{caleffi28} for $\zeta$ vs. $A^*$/$A$, and using parameters from Table~\ref{input} as input. We find $\zeta=224\pm 100$, with the uncertainty resulting from different parametrizations used to interpolate the data.  
\begin{table}
\setlength{\tabcolsep}{7pt}
\centering
\caption{Input parameters  for the calculation of the  concentrated loss coefficient in the distributors, $\zeta$ }
\label{input}
\begin{tabular}{@{} l l l @{}}
\hline\noalign{\smallskip}
Gas passage channel cross-section & $a^*$ &  0.00138 m$^2$\\ 
Number of gas passage channels per distributor & $n_P$ & $7$\\ 
Column cross-section & $A$  & $0.079$ m$^2$\\
$A^*$/$A$= $n_P\times a^*$/$A$&  &$0.12$ \\

\hline
\end{tabular}
\end{table}
Table~\ref{tab:dropcalc} shows the results obtained for the three runs, with the dominant uncertainty coming from the determination of $\zeta$.

\begin{table}
\setlength{\tabcolsep}{13pt}
\centering
\caption{Calculated velocity and density of the gas passing through the distributors, and  pressure drop due to the distributors  for runs~A,~B, and~C from Eq.~\ref{deltap}}
\begin{tabular}{@{} l c c c @{}}
\hline\noalign{\smallskip}
&$v_\mathrm{G}$(m s$^{-1}$)&$\rho_\mathrm{G}$(kg m$^{-3}$)&$\Delta p_{\mathrm{distributors}}$(mbar) \\
\noalign{\smallskip}\hline\noalign{\smallskip}
A &$0.202\pm 0.001$&$9.40\pm 0.01$&$1.7 \pm 0.7$\\ 
B &$0.235\pm 0.001$&$8.09\pm 0.01$&$2.0 \pm 0.9$\\ 
C &$0.242\pm 0.001$&$9.97\pm 0.01$&$2.6 \pm 1.1$\\ 
\hline
\end{tabular}
\label{tab:dropcalc}
\end{table}

\subsubsection{Calculation of the pressure drop in the packing}

The Sulcol and HYSYS programs are used to calculate the pressure drop per unit length of the packing for the dry bed, i.e. in the absence of liquid, $\Delta p_{\mathrm{packing,dry}}/\Delta z$ (the $z$ axis is directed downwards along the column). These calculations were cross-checked with a model for pressure drop calculation for packed columns described in  Ref.~\cite{stilch}, using the input parameters related to the packing reported in Table~14-14 of Ref.~\cite{Perry}.
The results are summarized in Table~\ref{dry} and show that  the three methods agree within~30\%.
The pressure drop in the packing increases with the liquid load since the liquid impedes the passage of gas.
\begin{table}
\setlength{\tabcolsep}{18pt}
\centering
\caption{Pressure drop per unit length of the packing for dry packing calculated for runs A, B, and C.}
\label{dry}
\begin{tabular}{@{} l l l l @{}}
\hline\noalign{\smallskip}
\multicolumn{4}{c}{{$\Delta p_{\mathrm{packing,dry}}/\Delta z$ (mbar m$^{-1}$)}} \\
\noalign{\smallskip}\hline\noalign{\smallskip}
& Sulcol& HYSYS& method of Ref.~\cite{stilch}\\

A &$0.45$&$0.49$&$0.38$\\ 
B &$0.52$&$0.56$&$0.44$\\  
C &$0.68$&$0.73$&$0.55$\\  
\hline
\end{tabular}
\end{table}

The authors of Ref.~\cite{stilch} also provide a method to estimate the pressure drop of the irrigated bed. 
Table~\ref{tabella2} summarizes the values obtained for the pressure drop per unit length of the irrigated bed, $\Delta p_{\mathrm{packing,irr}}/\Delta z$, which results about three times larger than that for the dry bed. We assigned an uncertainty of 30\% on this number based on the comparisons of Table~\ref{dry}.

\begin{table}
\setlength{\tabcolsep}{70pt}
\centering
\caption{Pressure drop per unit length for irrigated packing calculated for runs A, B and C using the method of Ref.~\cite{stilch}.}
\begin{tabular}{@{} l c @{}}
\hline\noalign{\smallskip}
 & {$\Delta p_{\mathrm{packing,irr}}/\Delta z$} (mbar m$^{-1}$)\\
\noalign{\smallskip}\hline\noalign{\smallskip}
A &$1.1\pm 0.3$\\ 
B &$1.2 \pm 0.4$\\ 
C &$1.6 \pm 0.4$\\ 
\hline
\end{tabular}
\label{tabella2}
\end{table}

The total pressure drop, $\Delta p_{\mathrm{tot,calc}}$, is given by the sum of the pressure drop of the distributors, from Table~\ref{tab:dropcalc}, and that of the irrigated packing, from Table~\ref{tabella2}  multiplied by the packing height of \AriaColumnModulePackingHeight, and is shown in Table \ref{tab:drop}.
\begin{table}
\setlength{\tabcolsep}{20pt}
\centering
\caption{Total experimental, $\Delta p_{\mathrm{tot,exp}}$,  and calculated, $\Delta p_{\mathrm{tot,calc}}$, pressure drop for tests A, B, and C.  }
\label{tab:drop}
\begin{tabular}{@{}  l c c @{}}
\hline\noalign{\smallskip}
  & $\Delta p_{\mathrm{tot,exp}}$ (mbar)& 
  $\Delta p_{\mathrm{tot,calc}}$ (mbar)\\
\hline\noalign{\smallskip}
A &$16.3\pm0.1$&$12.9\pm 4.0 $\\ 
B &$14.9\pm0.2$&$14.3\pm 4.5$\\  
C &$19.6\pm0.2$&$19.0\pm 6.0$\\  
\hline
\end{tabular}
\end{table}
Table~\ref{tab:drop} also shows the 
  measured  pressure drop of the column, $\Delta p_{\mathrm{tot,exp}}$,  i.e., the difference between the top and bottom recorded pressures (SIEMENS SITRANS P, DS III). 
Data and simulation show agreement within an uncertainty of the order of 30\%. We will assign a corresponding uncertainty to the values of $\Delta p_{\mathrm{packing,dry}}/\Delta z$, which will be used for comparison with Sulzer data in Fig.~\ref{gianni}.
 in relation  to  Table \ref{tab:drop} and Table \ref{dry}.

\subsection{Liquid holdup}
\label{sec:Hold-up}
The liquid holdup (or retention), $\mathrm{h_L}$, is the fraction of the packing volume occupied by the liquid. The holdup consists of two parts: static hold-up and dynamic hold-up. The dominant one is the dynamic hold-up, which
depends on the liquid load, and, to some extent, on the gas load. 

The holdup was evaluated using both HYSYS and Sulcol based on the operating conditions of the column, and with a direct calculation based on measured and estimated liquid volumes in the various elements of the plant, as described below.
Given the quantity of gas initially loaded, about $250$~kg, and the operating conditions of the column, we calculate that there are about $10$~kg of gas and $240$~kg of liquid, corresponding to $178$~L. The liquid is distributed both in the packed sections and in the processing circuit. There are about 80~L at the bottom of the column, as measured by a  level indicator, about 40~L in the reboiler, about 8~L in  the condenser (the  amount of liquid argon is related to the amount  of nitrogen in the  auxiliary circuit, as measured by a level indicator), about 11~L in the distributors (the height of the liquid level in the distributors during distillation must  be at least 3.5~cm, which is right above the height of the first hole in the distributor pipes), and about 2~L in the piping.
By subtraction, the amount of liquid in the packed sections is approximately 37~L, about~$20\%$ of the total liquid present in the column. 
 The stainless steel structure of the packing occupies about 15\% of the volume. 
 The estimated liquid holdup is, therefore, between~5.5\% and~6.0\%, as shown in Table~\ref{tab:hL}.  This relatively low value compares well with those obtained with HYSYS and Sulcol and agrees with measurements with the same packing by others~\cite{leveque}.
\begin{table}
\setlength{\tabcolsep}{20pt}
\centering
\caption{Liquid hold-up,  $\mathrm{h_L}$, calculations for runs A, B, and C.}
\label{tab:hL}
\begin{tabular}{@{} l l l l @{}}
\hline\noalign{\smallskip}
\multicolumn{4}{c}{\textbf{$\mathrm{h_L} (\%)$}} \\
\hline\noalign{\smallskip}
 &Sulcol& HYSYS & our estimate\\
A&$6.0$&$5.5$&$5.5$\\ 
B&$6.0$&$5.5$&$5.5$\\  
C&$6.5$&$5.5$&$6.0$\\  
\hline
\end{tabular}
\end{table}


%% file: sections/distillation.tex
\section{Measurement of distillation parameters}
\label{Results}

\begin{figure}
\centering

\includegraphics[width=\columnwidth]{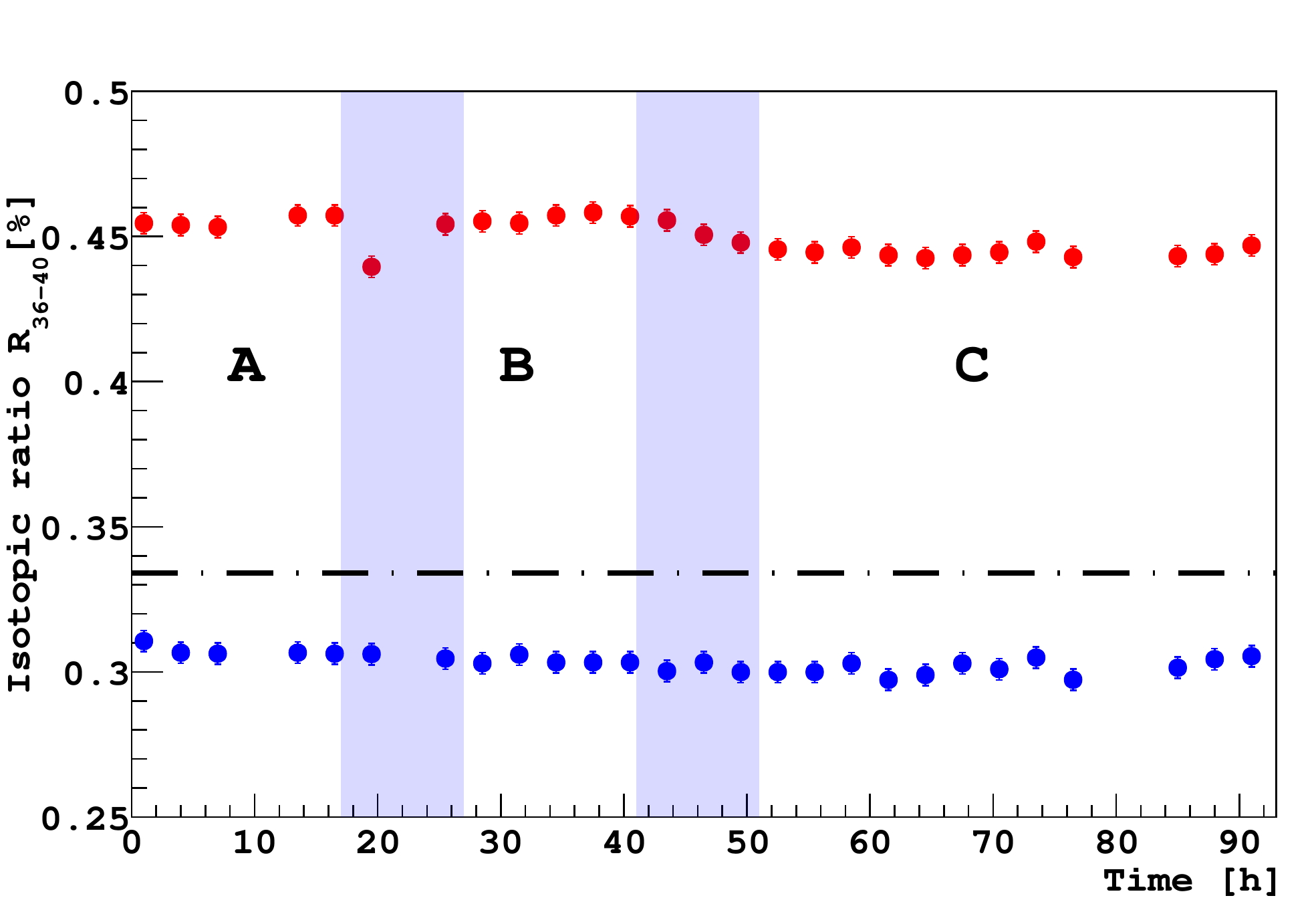}
\caption{The isotopic ratios $(R_{ 36-40})_T$ (red) and $(R_{ 36-40})_B$ (blue) as a function of time, after correction using the measured isotopic ratio in the feed, $(R_{36-40})_F$ (see text). The dashed-dotted line corresponds to the natural isotopic ratio. The shaded regions  correspond to changing conditions in the column (see Fig. \ref{totalissimo}) and are not included in  the data analysis. }
\label{soprasotto}
\end{figure}

As is well known, to achieve correct results from mass spectrometry in the measurements of isotopic ratios, it is necessary to get calibration factors. Therefore, we applied a time-dependent correction factor to the measured isotopic ratios $(R_{36-40})_T$ and $(R_{36-40})_B$, given by the ratio of the measured isotopic ratio in the feed, $(R_{36-40})_F$, and the known value of the natural isotopic ratio of 0.334\%.
Figure~\ref{soprasotto} shows the  corrected isotopic ratios $(R_{36-40})_T$ and $(R_{36-40})_B$ as a function of time.
The correction also compensates for time variations in the response of the  sampling system and reduces them by about a factor of ten.  
A systematic uncertainty on the isotopic ratios was evaluated, with reference to Fig.~\ref{spettro}, by comparing the  measurement of the isotopic ratio using the peak height value and the  integral of the mass distributions. We find a $\le 5$\% discrepancy.

\begin{figure}
\centering
\includegraphics[width=\columnwidth]{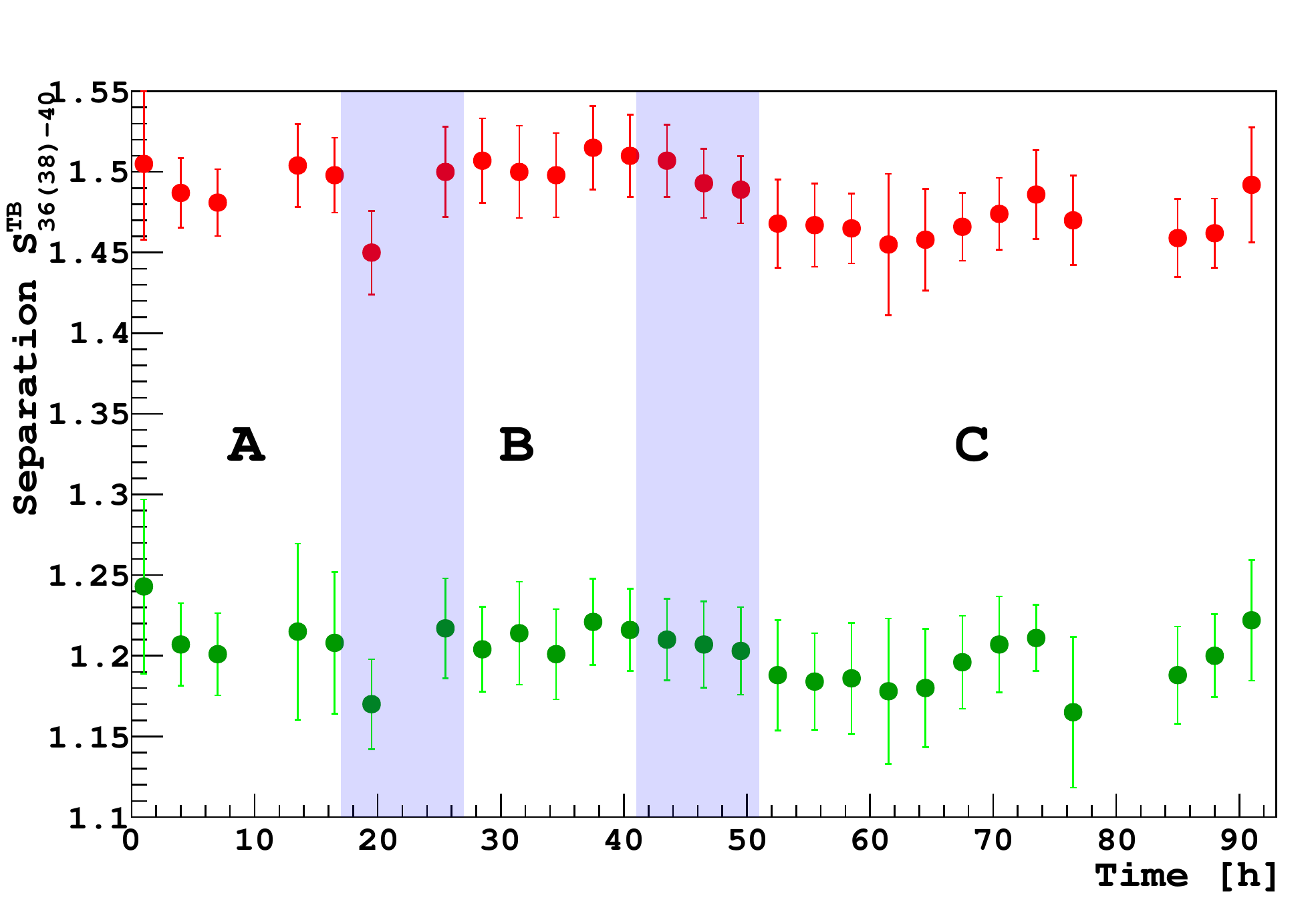}
\caption{Separation $S^{TB}_{36-40}$ (red) and $S^{TB}_{38-40}$ (green) as a function of time. The shaded regions  correspond to   changing conditions in the column (see Fig.~\ref{totalissimo}) and are not included in  the data analysis. }
\label{Setuptutti}
\end{figure}

The measured  separations $S^{TB}_{36-40}$ and $S^{TB}_{38-40}$ vs. time, defined in Eq.~\ref{frazione} are shown for the three runs~A, B, and C  in Fig.~\ref{Setuptutti}.
Time-dependent effects in this ratio of isotopic ratios largely cancel out and lead to negligible systematic uncertainty. The comparison of the measurement of separation values with the peak height value and the  integral of the mass distributions has a maximum discrepancy of about~2\%, of  the same size of  the statistical uncertainty.

We measure one separation value every three hours, with occasional data points not available due to connectivity problems with the mass spectrometer.
In runs~A and~B we observed a few~\% stability of $S^{TB}_{36-40}$, even though with  a limited number of measurements and duration of the run. In run~C $\sim~7$\% fluctuations are observed, possibly due to small pressure variations in the column during the run. We evaluated the impact of these fluctuations on $\ln{\alpha_{36-40}}$ in~\ref{vola} and found it negligible. Therefore, in the following, we use the weighted average of all measured separation values within a run.


The time-averaged separations for each run, ${\cal S}^{TB}_{36-40}$
and ${\cal S}^{TB}_{38-40}$, are shown in Table~\ref{tab1} and~\ref{tab2}.
\begin{table}
\centering
\caption{Time-averaged separation ${\cal S}^{TB}_{36-40}$  during runs A, B, and C,  logarithm of the relative volatility $\ln{\alpha_{36-40}}$,  equivalent number of theoretical stages $N$ and  HETP. }
\begin{tabular}{@{} l l l l l @{}}
\hline\noalign{\smallskip}
& ${\cal S}^{TB}_{36-40}$ &$\ln{\alpha_{36-40}} $  &  $N$ & HETP (cm)\\
\noalign{\smallskip}\hline\noalign{\smallskip}
A&$1.49  \pm0.03$&$(5.0  \pm 0.4)\times 10^{-3}$&$80  \pm 7$&$13  \pm 1 $ \\ 
B&$1.51  \pm 0.03$&$(5.2  \pm 0.4)\times 10^{-3}$&$79  \pm 7$&$13  \pm 1 $\\ 
C&$1.48  \pm 0.03$&$(4.9  \pm 0.4)\times 10^{-3}$&$80  \pm 8$&$13 \pm 1 $\\ 
\hline
\end{tabular}
\label{tab1}
\end{table}
\begin{table}
\centering
\caption{Time-averaged separation ${\cal S}^{TB}_{38-40}$  during runs A, B, and C,  logarithm of relative volatility $\ln{\alpha_{38-40}}$,  equivalent number of theoretical stages $N$ and  HETP.} 
\begin{tabular}{@{} l l l l l @{}}
\hline\noalign{\smallskip}
&  ${\cal S}^{TB}_{38-40}$& $\ln{\alpha_{38-40}}$ &  $N$ & HETP (cm)\\
\noalign{\smallskip}\hline\noalign{\smallskip}
A&$1.20 \pm 0.04$&$(2.4 \pm 0.3)\times 10^{-3}$&$76 \pm 17$&$14 \pm 3$\\ 
B&$1.21 \pm 0.03$&$(2.5 \pm 0.3)\times 10^{-3}$&$76 \pm 13$&$14 \pm 2$\\ 
C&$1.20\pm 0.03$ &$(2.3 \pm 0.3)\times 10^{-3}$&$79 \pm 15$&$13 \pm 2$\\ 
\hline
\end{tabular}
\label{tab2}
\end{table}
The corresponding relative volatility values were determined as discussed in~\ref{vola}, yielding an uncertainty of about 7\% for $\ln{\alpha_{36-40}}$ and about 12\% for $\ln{\alpha_{38-40}}$. The equivalent number of theoretical stages $N$ and  the HETP were calculated from  Eq.~\ref{eq1} and Eq.~\ref{eq3}. 
The values of $N$ and HETP obtained with the mass 36 and 38 isotopes and reported in the two tables agree within measurement uncertainties. Given their rather strong correlation, in the following, we only consider the measurement in Table~\ref{tab1}.

\begin{figure}
\centering
\includegraphics[width=1.05\columnwidth]{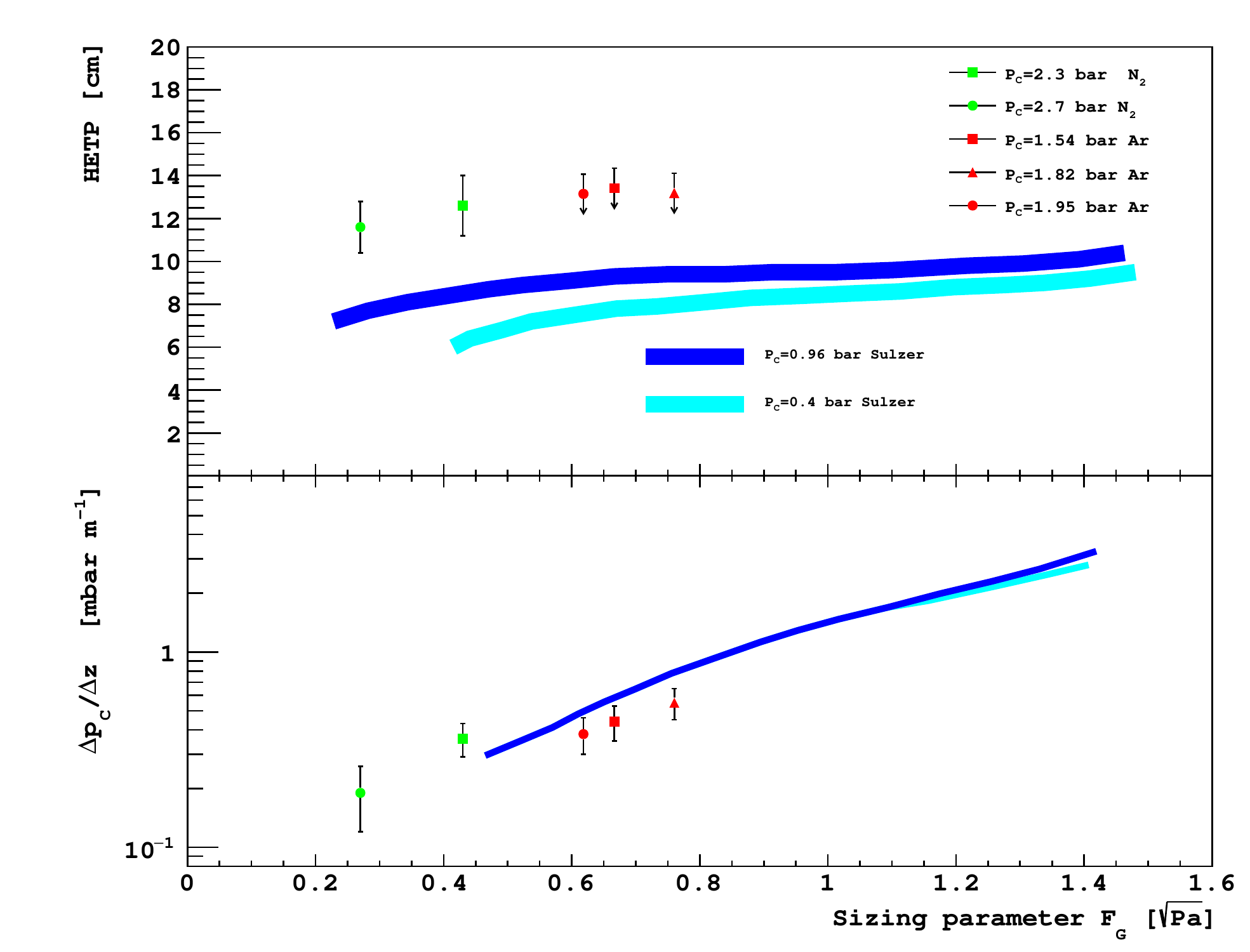}
\caption{
 Measured HETP (top), from Table~\ref{tab1} (red, green and violet dots), and  $\Delta p_{packing,dry}/\Delta z$ (bottom), from the third column of Table~\ref{dry} assuming 30\% uncertainty, vs.  sizing parameter, $F_G$, for the three runs with argon (red). Blue (cyan) line: measurements with chlorobenzene/ethylbenzene mixtures at $p_C=0.96$~bar ($0.40$~bar), Sulzer CY Gauze Packing, partial data-set extracted from the Structured Packing brochure of \href{https://www.sulzer.com/en/shared/products/gauze-packings}{Sulzer Ltd.}. A perfect match between our data and those from Sulzer is not expected due  to the different  thermodynamical parameters of the fluids. }
\label{gianni}
\end{figure}

Figure~\ref{gianni}  shows the measured HETP (top) from Table~\ref{tab1} and the calculated $\Delta p_{packing,dry}/\Delta z$ (bottom) (from Table~\ref{dry}) vs. the sizing parameter $F_G$. $F_G$ is the F-factor
or vapor load and is defined as $V_G$ $\times$ $\sqrt{\rho_G}$.

The three measured values of HETP are 20-30\% larger than the expectation based on  the Sulzer data. This is  possibly due to the different thermodynamical parameters between the argon and the fluid used by Sulzer. 

The measured separations and the derived HETP are conservative estimates due to the limited data taking time (from 15 h to  two days per run) at the same conditions and  the somewhat larger  average gas pressure in the column compared to the design value of 1.3~bar  presented  in Ref. \cite{Darkside:2021}.  

The measured pressure drop per unit length, which we assume for the Sulzer data to be the   pressure drop per unit length for the dry bed, follows quite well the  curve measured by Sulzer, within uncertainties.


%% file: sections/multi.tex
\section{{Multi-component distillation}}
\label{3otto}
In the paper on nitrogen distillation \cite{Darkside:2021},  we argued that 
the distillation of \ce{^39Ar} would not be affected by the presence of the more abundant \ce{^36Ar} and \ce{^38Ar}, based on the argument that for a gas mixture with one dominant component and several other components with mass fractions below a few 0.1\% each, the distillation of these is essentially independent of the presence of the others and is well described as a binary distillation with respect to \ce{^40Ar}. Therefore, a multi-component approach for distillation of argon~\cite{Kister:1992} is not needed. This argument does not hold, e.g., for xenon~\cite{Back}.
This assumption was directly tested with this run. 

In total reflux condition, if the distillation \ce{^36Ar} and \ce{^38Ar} proceed independently of each other, we expect

\begin{equation} S^{TB}_{36-40}=(\alpha_{36-40})^{N}
\end{equation}

and 

\begin{equation} S^{TB}_{38-40}=(\alpha_{38-40})^{N}
\end{equation}


From Eq.\ref{dici} one infers that:



\begin{equation} \label{kappaf} k_{36-38}\coloneqq \frac{\mathrm{ln}(S^{TB}_{36-40})}{\mathrm{ln}(S^{TB}_{38-40})}\times \frac{1}{2.1}=1. \end{equation}

\begin{figure}
\centering
\includegraphics[width=\columnwidth]{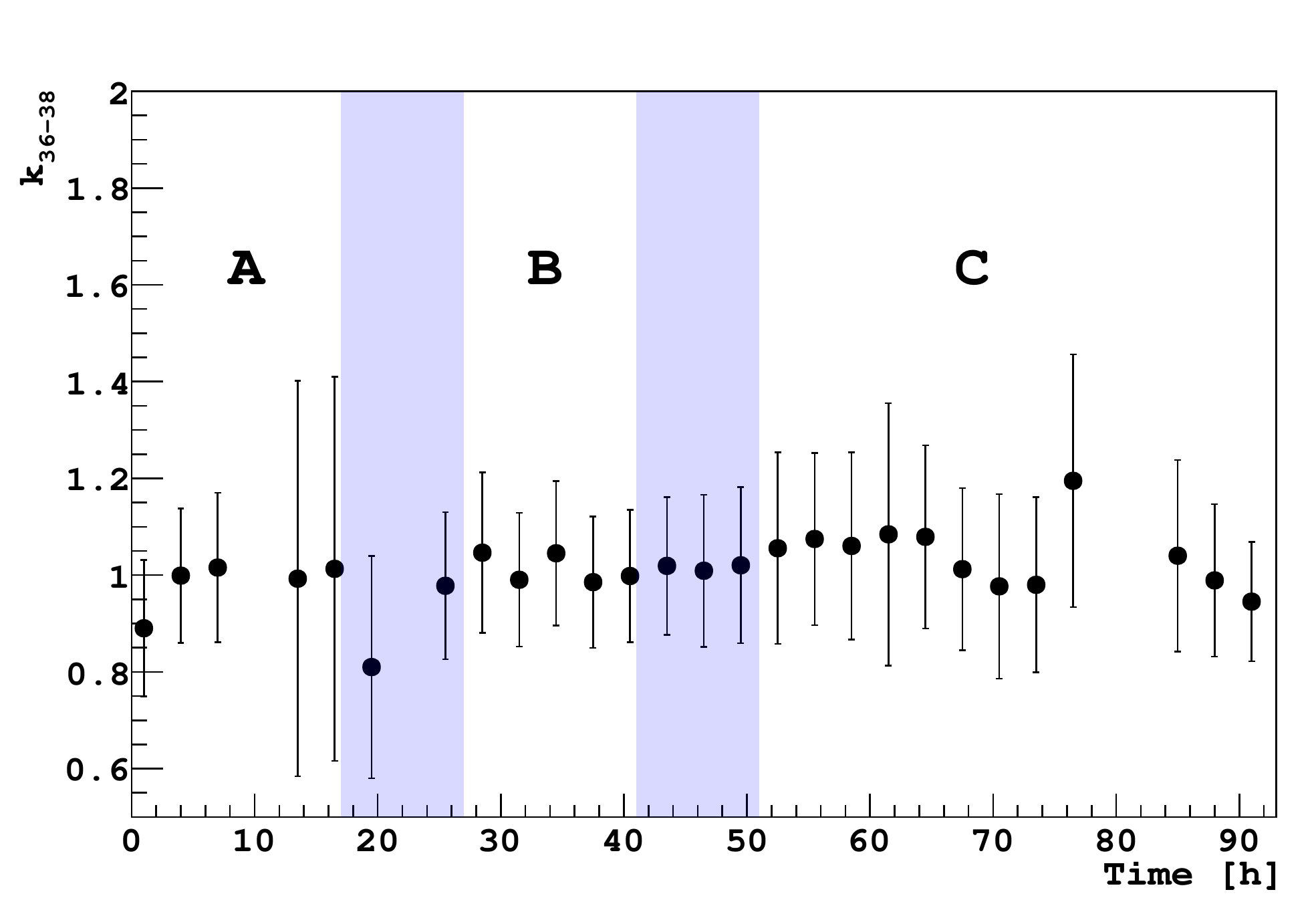}
\caption{Measured $k_{36-38}$  (as defined in Eq. \ref{kappaf}) as a function of time. The shaded regions  correspond to   changing conditions in the column (see Fig.~\ref{totalissimo}) and are not included in  the data analysis.}
\label{Setup5}
\end{figure}
Fig.~\ref{Setup5} shows the measured value of $k_{36-38}$ 
vs time. The time average of these values is $0.998 \pm 0.030$, consistent with 1, and demonstrates that no  multi-component approach is needed for argon distillation.

%% file: sections/comparison.tex
\section{Comparison with simulation}
\label{sec:comp}

The measured distillation data for run~B were compared to simulations with HYSYS.
We used the conditions of run~B as input  parameters for the simulation. These are the mean values of $p$ from Table~\ref{tab_dati}, $V$ from Table~\ref{tab_altridati}, HETP from Table~\ref{tab1}, and ${\cal S}^{FB}_{36-40}$ from~Table~\ref{confronto}.
The comparison of measured and simulated separations is shown in the first two columns of  Table~\ref{confronto}. Good agreement is observed within the uncertainties. 
The simulation was also performed for 
 binary systems using both the rigorous and the MCT methods.  The latter is included for completeness and because of its use in Ref.~\cite{Darkside:2021}. We find good agreement across the board.
\begin{table}
\setlength{\tabcolsep}{10pt}
\centering
\caption{Comparison of separations between measurements and simulation, related to run B conditions: data (Exp.) from Table \ref{tab1} and Table \ref{tab2}, HYSYS rigorous simulation 
with three components (Rig.~3), 
with two components (Rig.~2) and      McCabe-Thiele (MCT) simulation.}
\begin{tabular}{@{} l l l l l @{}}
\hline\noalign{\smallskip}
& Exp. & Rig.~3 & Rig.~2 & MCT \\
\noalign{\smallskip}\hline\noalign{\smallskip}
${\cal S}^{TB}_{36-40}$& {$1.51 \pm 0.03$}  &$1.5 \pm 0.1$& $1.5 \pm 0.1 $ &$1.5 \pm 0.1$\\ 
${\cal S}^{TB}_{38-40}$& {$1.21 \pm 0.03$} & $1.20 \pm 0.04$ & \\ 
\hline
\end{tabular}
\label{confronto}
\end{table}



%% file: sections/conclusions.tex
\section{Conclusions and outlook}
\label{sec:conclusions}

We validated the performance of the Aria plant with argon running the prototype plant for a few days in total reflux. 
We significantly improved our understanding of the plant performance compared to the previous nitrogen distillation run, achieving a deeper understanding of operating characteristics such as the hydraulic parameters and the behavior of multi-component distillation.
We also successfully modeled the Aria prototype performance using commercial process simulation software. This study is a validation of both the plant behavior and the software itself. Originally designed for distilling organic compounds at room temperature, it is also suitable for the simulation of cryogenic isotopic distillation.

The average measured HETP is about 30\% larger than that measured in a different column with the same packing, using organic mixtures at room temperature by Sulzer. The discrepancy could be due to the different thermodynamical parameters of the fluids used in the two cases or to 
the somewhat larger  average gas pressure in the column compared to the design value of 1.3~bar  presented  in Ref. \cite{Darkside:2021}. 

The measured pressure drop per unit length is in agreement, within uncertainties, with that  measured with the organic mixture. 



%% file: sections/Acknoledgment.tex
\section*{Acknowledgements}


This report is based upon work supported by FSC 2014-2020 - Patto per lo Sviluppo, Regione Sardegna, Italy, the U. S. National Science Foundation (NSF) (Grants No. PHY-0919363, No. PHY-1004054, No. PHY-1004072, No. PHY-1242585, No. PHY-1314483, No. PHY- 1314507, associated collaborative grants, No. PHY-1211308, No. PHY-1314501, and No. PHY-1455351, as well as Major Research Instrumentation Grant No. MRI-1429544), the Italian Istituto Na\-zionale di Fisica Nucleare (Grants from Italian Ministero dell’Istruzione, Università, e Ricerca Progetto Premiale 2013 and Commissione Scientific Nazionale II), the Natural Sciences and Engineering Research Council of Canada,
SNOLAB, and the Arthur B. McDonald Canadian Astroparticle Physics Research Institute. 
We acknowledge the financial support by LabEx UnivEarthS (ANR-10-LABX-0023 and ANR18-IDEX-0001), the São Paulo Research Foundation (Grant FAPESP-2017/26238-4), Chinese Academy of Sciences (113111KYSB20210030) and National Natural Science Foundation of China (12020101004).
The authors were also supported by the Spanish Ministry of Science and Innovation (MICINN) through the grant PID2019-109374GB-I00, the ``Atraccion de Talento'' Grant 2018-T2/\ TIC-10494, 
the Polish NCN, Grant No. UMO-\ 2019/\ 33/\ B/\ ST2/\ 02884, the Polish Ministry of Science and Higher Education, MNi\-SW, grant number 6811/IA/SP/2018, the International Research Agenda Programme AstroCeNT, Grant No. MAB\-/2018/7, funded by the Foundation for Polish Science from the European Regional Development Fund, the European Union’s Horizon 2020 research and innovation program under grant agreement No 952480 (DarkWave), the Science and Technology Facilities Council, part of the United Kingdom Research and Innovation, and The Royal Society (Uni\-ted Kingdom), and IN2P3-COPIN consortium (Grant No. 20-152).  
I.F.M.A is supported in part by Conselho Nacional de Desenvolvimento Científico e Tecnológico (CNPq). We also wish to acknowledge the support from Pacific Northwest National Laboratory, which is operated by Battelle for the U.S. Department of Energy under Contract No. DE-\-AC05-76RL01830.
This research was supported by the Fer\-mi National Accelerator Laboratory (Fermilab), a U.S. Department of Energy, Office of Science, HEP User Facility. Fermilab is managed by Fermi Research Alliance, LLC \- (FRA), acting under Contract No. DE-AC02-07CH11359.

We acknowledge the   professional contribution of  the Mi\-ne and  Electrical Maintenance staff of Carbosulcis S.p.A. to this activity.
We thank Polaris S.r.L. and in particular M. Masetto and E.V.Canesi for their continuous support during the preparation and execution of the run. We thank  Fondazione Aria for its contribution to the project and, in particular, for the help during the data-taking. We thank S.Nisi of Istituto Nazionale di Fisica Nucleare, Laboratori Nazionali del Gran Sasso, for his contribution in the commissioning of  the sampling system and M.Guetti of Istituto Nazionale di Fisica Nucleare, Laboratori Nazionali del Gran Sasso, for the help in  plant maintenance.
We thank M. Arba and M. Tuveri of the Cagliari Division of Istituto Nazionale di Fisica Nucleare for their support during the preparation of the run.
We acknowledge the contribution of eng. S. Tosti, T. Pinna, D.Dongiovanni, A.Santucci of ENEA for the safety analyses for Aria.

%% file: sections/vola.tex
\section{Evaluation of relative volatilities}
\label{vola}
In this section, we give an update of our best estimate of the  relative volatilities $\alpha_{36-40}$ and $\alpha_{38-40}$, based on the scrutiny of  the literature, and we update  the result derived in Appendix A of Ref. \cite{Darkside:2021}.
For $\mathrm{ln\alpha_{36-40}}$, we included   measurements from Ref.~\cite{Bigeleisen:1972} and  from Ref.~\cite{Lee:1970a}, corresponding to the second definition of  relative volatility of Eq.~\ref{nuova}.

\begin{figure}
\centering
\includegraphics[width=\columnwidth]{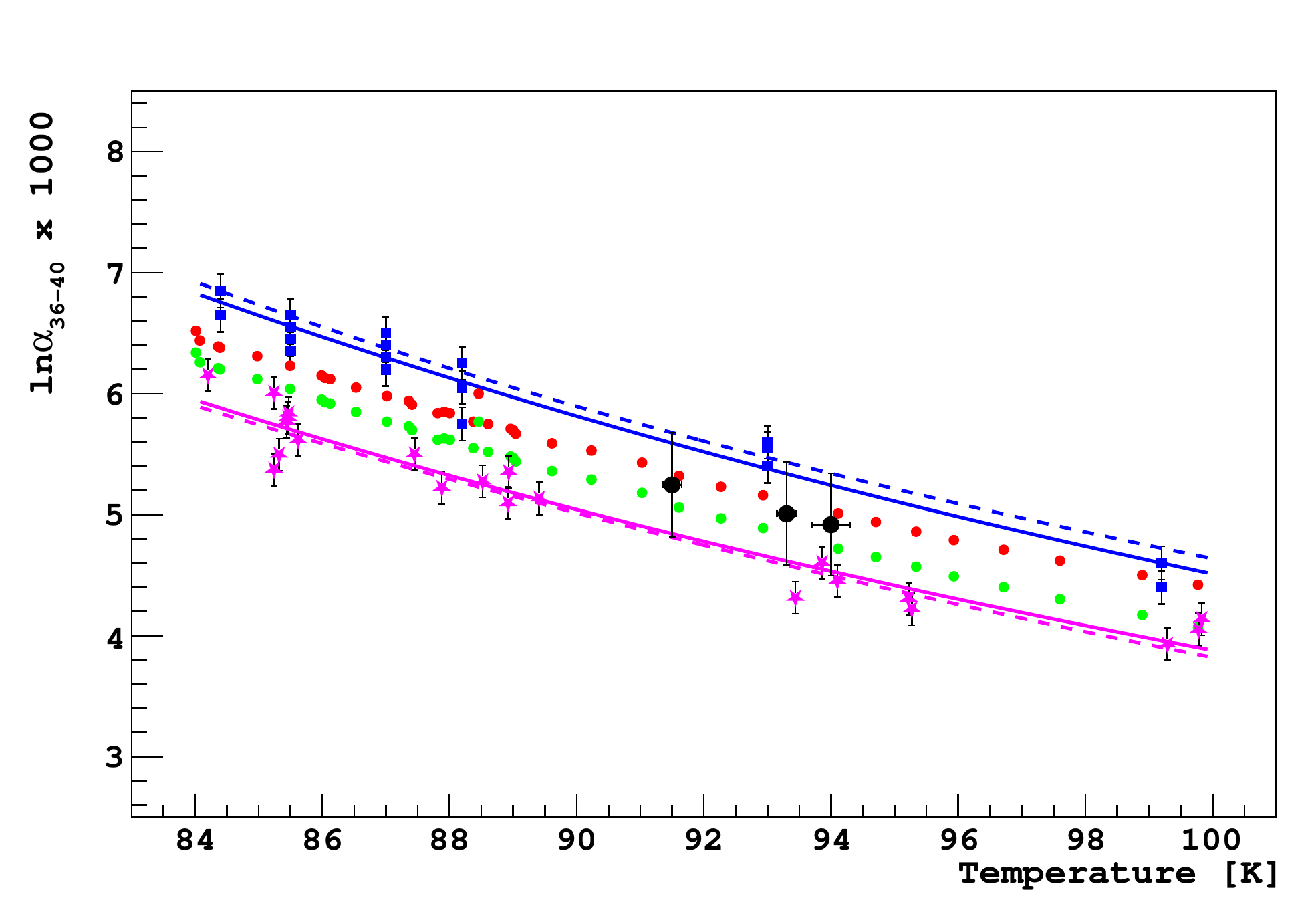}
\caption{Measurements of  $\mathrm{ln\alpha_{36-40}}$ vs. temperature. The data  are taken from~\cite{Boato1961} (blue squares), \cite{Bigeleisen:1972} (violet stars), and~\cite{Lee:1970a} (green and red dots).   The function $\mathrm{ln{\alpha}_{36-40}} = A\times 1/T^2 + B $  is fitted (continuous lines)  to the data of Ref.~\cite{Boato1961} and Ref.~\cite{Bigeleisen:1972}. The corresponding one-sided standard deviation curves (dashed lines)are also shown.  The derived values of $\mathrm{ln\alpha_{36-40}}$ corresponding to the  three runs A, B, and C are also shown (black dots).
}
\label{vola1}
\end{figure}

In Fig.~\ref{vola1} we report the measured dependence of $\mathrm{ln\alpha_{36-40}}$ on temperature.
The function $\mathrm{ln{\alpha}_{36-40}} = A\times 1/T^2 + B $  is fitted to the data of Ref. \cite{Boato1961} and Ref. \cite{Bigeleisen:1972}.
The choice of this  parametrization follows the theoretical considerations of Ref.~\cite{Boato1962b}. 
The errors on the single measurement  were set all equal in the fit and determined in retrospect  requiring the reduced $\chi^2$ to be one.
Applying  error propagation  for the estimate of  $\mathrm{ln{\hat{\alpha}}_{36-40}} = \hat{A}\times 1/T^2 + \hat{B} $ and of the uncertainty 
$\sigma_{\mathrm{ln\hat{\alpha}}}$ as:
\begin{equation}
\sigma_{\mathrm{ln\hat{\alpha}}}=\sqrt{V_{00}\times (1/T^2)^2+ V_{11} + 2 /T^2 \times V_{22}}
\end{equation}
with $V_{ij}$ being the elements of the covariance matrix, one obtains  the mean (continuous lines) and one-sided standard deviation curves (dashed lines). 
To take into account all existing measurements in our estimate, we assumed that the one standard deviation value  of $\mathrm{ln\alpha_{36-40}}$ at a given temperature lies inside the two dashed curves.
The values of $\mathrm{ln\alpha_{36-40}}$ corresponding to the temperatures of the three runs A, B, and C are also shown and their values are reported in Table~\ref{tab1}. The horizontal error bars correspond to the uncertainty on the temperature related to the pressure variations during the runs of Table~\ref{tab_dati}. They give  a negligible contribution to the uncertainty on $\mathrm{ln\alpha_{36-40}}$. 


The relative volatility $\ln{\alpha_{38-40}}$, at given temperature,  was derived from $\ln {\alpha_{36-40}}$ using the theoretical formula   from Ref.~\cite{CanongiaLopes:2003ju}, Eq.~(25):

\begin{equation}
\ln{\alpha_{A-40}}\propto \frac{40-A}{A^3},
\end{equation}

with $A$ the atomic mass of an isotope, which entails:

\begin{equation} \label{dici} \ln{\alpha_{36-40}}\sim \ln{\alpha_{38-40}} \times 2.1.\end{equation}

This dependence is also consistent, within errors, with the measurements of Ref.~\cite{Alamre:2020}, of  $\ln{\alpha_{36-40}}$ = \- $\ln{\alpha_{38-40}}$ $\times (2.3 \pm 0.2)$ 
at $87.3$ K. The relative uncertainty from these measurements was propagated on the final estimate of the quantity $\ln{\alpha_{38-40}}$.



%% file: basic/authors_ds20k-copy.tex
\newcommand{\notds}{\nolinebreak\footnotemark\nolinebreak}
\renewcommand{\thefootnote}{$*$}

\onecolumn
\textbf{The DarkSide-20k Collaboration}
\footnotetext{Not a member of the DarkSide-20k Collaboration}

E.~Aaron\thanksref{UCDavis}\nolinebreak,
P.~Agnes\thanksref{AQGSSI}\nolinebreak,
I.~Ahmad\thanksref{AstroCeNT}\nolinebreak,
S.~Albergo\thanksref{CTINFN}{CTUNI}\nolinebreak,
I.~F.~M.~Albuquerque\thanksref{USP}\nolinebreak,
T.~Alexander\thanksref{PNNLaddress}\nolinebreak,
A.~K.~Alton\thanksref{Augustana}\nolinebreak,
P.~Amaudruz\thanksref{TRIUMFaddress}\nolinebreak,
M.~Atzori Corona\thanksref{CAUniPHY}{CAINFN}\nolinebreak,
M.~Ave\thanksref{USP}\nolinebreak
% ...

\begin{enumerate}[label=\textsuperscript{\arabic*}]
    \item {\label{Houston}\Houston}
    \item {\CTINFN \label{CTINFN}}
    \item {\CTUNI \label{CTUNI}}
    \item {\USP \label{USP}}
    \item {\PNNLaddress \label{PNNLaddress}}
    \item {\BOUniPHY \label{BOUniPHY}}
    \item {\BOINFN \label{BOINFN}}
    \item {\Augustana \label{Augustana}}
    \item {\TRIUMFaddress \label{TRIUMFaddress}}
    \item {\MendeleevUniverisity \label{MendeleevUniverisity}}
    \item {\LNLINFN \label{LNLINFN}}
    \item {\RHUL \label{RHUL}}
    \item {\MSU \label{MSU}}
    \item {\CIEMAT \label{CIEMAT}}
    \item {\CPPM \label{CPPM}}
    \item {\NAINFN \label{NAINFN}}
    \item {\PIINFN \label{PIINFN}}
    \item {\PIUniPHY \label{PIUniPHY}}
    \item {\BINP \label{BINP}}
    \item {\NSU \label{NSU}}
    \item {\CAINFN \label{CAINFN}}
    \item {\GEUni \label{GEUni}}
    \item {\GEINFN \label{GEINFN}}
    \item {\Carleton \label{Carleton}}
    \item {\CERNaddress \label{CERNaddress}}
    \item {\RMTreINFN \label{RMTreINFN}}
    \item {\RMTreUni \label{RMTreUni}}
    \item {\CAUniPHY \label{CAUniPHY}}
    \item {\AQLNGS \label{AQLNGS}}
    \item {\AQGSSI \label{AQGSSI}}
    \item {\CentroFermi \label{CentroFermi}}
    \item {\CAUniEEE \label{CAUniEEE}}
    \item {\NAUniPHY \label{NAUniPHY}}
    \item {\NAUniStruct \label{NAUniStruct}}
    \item {\OACINAF \label{OACINAF}}
    \item {\Zaragoza \label{Zaragoza}}
    \item {\Krakow \label{Krakow}}
    \item {\MIINFN \label{MIINFN}}
    \item {\TOINFN \label{TOINFN}}
    \item {\LPNHE \label{LPNHE}}
    \item {\RMUnoINFN \label{RMUnoINFN}}
    \item {\RMUnoUni \label{RMUnoUni}}
    \item {\SAUni \label{SAUni}}
    \item {\SAINFN \label{SAINFN}}
    \item {\MIPoliCHE \label{MIPoliCHE}}
    \item {\Petersburg \label{Petersburg}}
    \item {\Kurchatov \label{Kurchatov}}
    \item {\UMass \label{UMass}}
    \item {\UCDavis \label{UCDavis}}
    \item {\LNFINFN \label{LNFINFN}}
    \item {\APC \label{APC}}
    \item {\Princeton \label{Princeton}}
    \item {\ETHZ \label{ETHZ}}
    \item {\TOPoli \label{TOPoli}}
    \item {\WilliamsCollege \label{WilliamsCollege}}
    \item {\Hawaii \label{Hawaii}}
    \item {\TNFBK \label{TNFBK}}
    \item {\TNTIFPA \label{TNTIFPA}}
    \item {\UB \label{UB}}
    \item {\MEPhI \label{MEPhI}}
    \item {\JINR \label{JINR}}
    \item {\IHEPaddress \label{IHEPaddress}}
    \item {\ENUniCEE \label{ENUniCEE}}
    \item {\CTLNS \label{CTLNS}}
    \item {\Alberta \label{Alberta}}
    \item {\AstroCeNT \label{AstroCeNT}}
    \item {\FNALaddress \label{FNALaddress}}
    \item {\SNOLABaddress \label{SNOLABaddress}}
    \item {\Laurentian \label{Laurentian}}
    \item {\Belgorod \label{Belgorod}}
    \item {\NAUniPHARM \label{NAUniPHARM}}
    \item {\ZaragozaARAID \label{ZaragozaARAID}}
    \item {\Queens \label{Queens}}
    \item {\Lancaster \label{Lancaster}}
    \item {\MIPoliICA \label{MIPoliICA}}
    \item {\Manchester \label{Manchester}}
    \item {\MIUni \label{MIUni}}
    \item {\BNLaddress \label{BNLaddress}}
    \item {\NAUniCHE \label{NAUniCHE}}
    \item {\VTech \label{VTech}}
    \item {\UCLA \label{UCLA}}
    \item {\FortLewis \label{FortLewis}}
    \item {\Lodz \label{Lodz}}
    \item {\CAUniCHE \label{CAUniCHE}}
    \item {\UCAS \label{UCAS}}
    \item {\Mainz \label{Mainz}}
    \item {\UCRiverside \label{UCRiverside}}
    \item {\Liverpool \label{Liverpool}}
    \item {\Birmingham \label{Birmingham}}
    \item {\STFCInterconnect \label{STFCInterconnect}}
    \item {\Temple \label{Temple}}
    \item {\UniversityofEdinburgh \label{UniversityofEdinburgh}}
    \item {\Warwick \label{Warwick}}
    \item {\UnivAQ \label{UnivAQ}}
     \item {\CS \label{CS}}
    \end{enumerate}